\documentclass{article}
\usepackage{graphicx} % Required for inserting images

\usepackage[margin=1in]{geometry} % full-width

\usepackage{amsmath}
\usepackage{amsthm}
\usepackage{amssymb}
\usepackage{overpic}
\usepackage{comment}

\usepackage[utf8]{inputenc}
\usepackage[hidelinks]{hyperref}
\hypersetup{
	unicode,
	pdfauthor={Author One, Author Two, Author Three},
	pdftitle={A simple article template},
	pdfsubject={A simple article template},
	pdfkeywords={article, template, simple},
	pdfproducer={LaTeX},
	pdfcreator={pdflatex}
}

\usepackage[sort&compress,numbers,square]{natbib}
\numberwithin{equation}{section} 
\theoremstyle{plain}
\newtheorem{theorem}{Theorem}

\newtheorem{proposition}[theorem]{Proposition}

\theoremstyle{definition}

\usepackage{graphicx, color}
\graphicspath{{fig/}}

\usepackage{comment}
\usepackage{color}

\usepackage{algorithm, algpseudocode} 
\usepackage{mathrsfs} 

\usepackage{lipsum}

\def\be{\begin{equation}}
	\def\ee{\end{equation}}
\def\bse{\begin{subequations}}
	\def\ese{\end{subequations}}

\def\gl{\mathrel{\mathpalette\overl@ss>}}

\def\sech{\mathop{\rm sech}\nolimits}

\def\Real{\mathbb{R}}

\def\Complex{\mathbb{C}}

\def\@#1{{\mathbf{#1}}}
\def\_#1{{\mathsf{#1}}}
\let\~=\tilde
\let\==\bar
\let\^=\hat
\let\<=\langle
\let\>=\rangle

\def\note[#1]{\marginpar{\color{red}[#1]}}

\def\XXint#1#2#3{{\setbox0=\hbox{$#1{#2#3}{\int}$}
		\vcenter{\hbox{$#2#3$}}\kern-.5\wd0}}

\def\1{{\bf 1}}

\makeatother
\usepackage{color}

\newcommand{\CC}{\mathbb{C}}

\renewcommand{\1}{{\bf 1}}
\newcommand{\lda}{\lambda}

\title{
Breather interactions in the integrable discrete Manakov system and trigonometric Yang-Baxter maps}
\author{V. Caudrelier$^1$, N. J. Ossi$^{2}$, and B. Prinari$^{2}$}

\date{
$^1$ School of Mathematics, University of Leeds, Leeds, UK \\
$^2$ Department of Mathematics, State University of New York, Buffalo, NY 14260 USA\\
}

\begin{document}
	\maketitle
%%%%%%%%%%%%%%%%%%%%%%%%%%%%%%%%%%%%%%%%%%%%%%%%%%%	
\begin{abstract}
The goal of this work is to obtain a 
%{\color{red}
complete 
%(?)} 
characterization of soliton and breather interactions in the integrable discrete Manakov (IDM) system, a vector generalization of the Ablowitz-Ladik model.
%(which, in turn, is an integrable discretization of the nonlinear Schr\"odinger (NLS) equation). 
The IDM system, which in the continuous limit reduces to the Manakov system (i.e., a 2-component vector nonlinear Schr\"odinger equation), was shown to admit a variety of discrete vector soliton solutions: fundamental solitons, fundamental breathers, and composite breathers. While the interaction of fundamental solitons was studied early on, no results are presently available for other types of soliton-breather and breather-breather interactions. Our study reveals that upon interacting with a fundamental breather, a fundamental soliton
becomes a fundamental breather. Conversely, the interaction of two fundamental breathers generically yields two fundamental breathers with polarization shifts, but may also result in a fundamental
soliton and a fundamental breather. Composite breathers interact trivially both with each other and with a fundamental soliton or breather. Explicit formulas for the scattering coefficients that characterize fundamental and composite 
breathers are given. This allows us to interpret the interactions in terms of a refactorization problem and derive the associated Yang-Baxter maps describing the effect of interactions on the polarizations. These give the first examples of parametric Yang-Baxter maps of trigonometric type.
\end{abstract}
%%%%%%%%%%%%%%%%%%%%%%%%%%%%%%%%%%%%%%%%%%%%%%%%%%%
%\tableofcontents
%%%%%%%%%%%%%%%%%%%%%%%%%%%%%%%%%%%%%%%%%%%%%%%%%%%
	\section{Introduction}
%The goal of this work is to characterize soliton and breather interactions for 
In this work we consider the following system of differential-difference equations:
\begin{equation}
\label{e:idMq}
i\frac{d \@q_n}{dt}=\frac{1}{h^2}(\@q_{n+1}-2\@q_n+\@q_{n-1})-\sigma\|\@q_n\|^2(\@q_{n+1}+\@q_{n-1}),
\qquad \sigma=\mp 1,
\end{equation}
where $\@q_n(t)$ is a 2-component complex vector function of $n\in \mathbb{Z}$, $t\in\Real$, and $\sigma=\mp 1$ distinguishes between the focusing/defocusing dispersion regimes.
This system, which we will refer to as the integrable discrete Manakov (IDM) system, was introduced in \cite{GI81a,GI81b,GI82} as 
a vector generalization of the Ablowitz-Ladik model (Eq.~\eqref{e:idMq} for a scalar $q_n(t)$, see \cite{AL1,AL2}), and it is an 
integrable $\mathcal{O}(h^2)$ spatial discretization of the 
%{\color{blue} focusing (keep or remove $\sigma$?)} 
Manakov system \cite{M74}:
$$
i\@q_t=\@q_{xx}-2\sigma \|\@q\|^2\@q, \qquad \sigma=\mp 1,
$$
to which it reduces as the lattice spacing $h\to 0$ with $nh\to x$. In turn, the Manakov system is a vector generalization of the celebrated nonlinear Schr\"odinger (NLS) equation \cite{ZS72}, and, like its scalar counterpart, it is a completely integrable system. In particular, both the Manakov system and its integrable discretization \eqref{e:idMq} are linearizable by the Inverse Scattering Transform (IST), and they admit vector soliton solutions. In this work we will consider the focusing case, so we will take $\sigma=-1$ throughout.
The vector solitons of the focusing Manakov system have the form:
\begin{equation}
\label{e:1solitonM}
\@q(x,t)= q(x,t)\, \@p, \qquad q(x,t)=2\eta e^{-2i\xi x+4i(\xi^2-\eta^2)t}\sech(2\eta x-8\xi t-2x_o),
\end{equation}
where $q(x,t)$ is the 1-soliton solution of the scalar NLS equation, and $\@p\in \Complex^2$ is a norm-1 complex vector. From a spectral point of view, the physical parameters that characterize the soliton are encoded in: a ``discrete eigenvalue'' $k=\xi+i\eta\in \Complex^+$, whose real and imaginary parts, respectively, fix the soliton velocity ($v=4\xi$) and amplitude ($A=2\eta>0$);  and a ``norming constant'' $\@C\in \Complex^2$ associated to the discrete eigenvalue $k$, which determines the center of the soliton, $x_o=\log \sqrt{||\@C||/2\eta}$, and its ``polarization'' $\@p=\@C^\dagger/||\@C||$.

While a 1-soliton solution of the Manakov system is fundamentally governed by the scalar NLS, the vector nature of the solution affects the dynamics when solitons with different polarizations interact. Indeed, as was already established in Manakov's pioneering paper \cite{M74}, interacting vector solitons generically change their polarizations upon interaction. Specifically, for a 2-soliton solution with individual solitons traveling at different velocities, in the backward ($t\to -\infty$) and forward ($t\to+\infty$) long-time limits such a solution asymptotically breaks up into individual solitons
\begin{equation}
\@q(x,t)\sim\@q^\pm(x,t)=\@p_1^\pm q_1^\pm(x,t)+\@p_2^\pm q_2^\pm(x,t) \qquad \text{as }t\to \pm \infty,
\end{equation}
where $\@p_j^\pm$ are complex unit vectors, and $q_j^\pm(x,t)$ are 1-soliton solutions of the scalar NLS such that $q_j^-$ and $q_j^+$ are characterized by the same amplitude and velocities as determined by the discrete eigenvalues $k_j=\xi_j+i\eta_j$ for $j=1,2$, with $\xi_1\ne \xi_2$. Manakov's formulas express the polarizations of the solitons in the forward long-time limit $\@p_1^+,\@p_2^+$ in terms of the polarizations $\@p_1^-,\@p_2^-$:
\bse
\label{e:Manakov_p}
\begin{gather}
\@p_2^+=\frac{1}{\chi}\frac{k_1-k_2^*}{k_1^*-k_2^*}
\left[ \@p_2^-+\frac{k_1^*-k_1}{k_2^*-k_1^*}(\@p_1^{-\, \dagger}\@p_2^-)\@p_1^-\right], \\
\@p_1^+=\frac{1}{\chi}\frac{k_1-k_2^*}{k_1-k_2}
\left[ \@p_1^-+\frac{k_2^*-k_2}{k_2-k_1}(\@p_2^{-\, \dagger}\@p_1^-)\@p_2^-\right],
\end{gather}
with
\begin{gather}
\chi^2=\left|\frac{k_1-k_2^*}{k_1-k_2} \right|^2
\left[ 1+\frac{(k_1^*-k_1)(k_2-k_2^*)}{|k_1-k_2|^2}
|\@p_1^{-\, \dagger} \@p_2^-|^2\right],
\end{gather}
\ese
while the centers of the solitons in the forward-backward long-time limits are given by:
\begin{equation}
e^{2(x_2^+-x_2^-)}=\chi\,, \qquad e^{2(x_1^+-x_1^-)}=1/\chi\,.
\end{equation}
[Here and in the following: $^*$ denotes complex conjugation, and the superscripts $^T$ and $^\dagger$ are used for matrix transpose and conjugate transpose, respectively].
The Manakov formulas \eqref{e:Manakov_p} show that in general when the solitons interact, the intensity distributions of the individual solitons in the two components change, yielding a ``polarization shift'', and only when the ``initial'' polarizations of the solitons $\@p_1^-,\@p_2^-$ are either parallel or orthogonal is the amplitude of each individual component of the solitons conserved. 
Eqs.~\eqref{e:Manakov_p} were obtained in \cite{M74} by tracing the asymptotic states of the eigenfunctions 
%of the scattering problem associated to the Manakov system \eqref{e:Manakov} 
through each soliton, multiplying them by the corresponding soliton transmission coefficient, and accounting for bound states of the eigenfunctions via the norming constants.
In the multisoliton case, a $J$ soliton collision is equivalent to the composition of $J(J-1)/2$ pairwise interactions taking place in an arbitrary order compatible with the soliton velocities, and the effect of the interactions (i.e., the shifts in the soliton centers and in the polarizations of each individual soliton) is independent of the order in which the pairwise solitons actually interact \cite{APT04a}. In turn, this was shown to be related to the fact that the map $R[k_1,k_2]:(\@p_1^-\,,\,\@p_2^-)\mapsto (\@p_1^+\,,\,\@p_2^+)$ defined by   \eqref{e:Manakov_p} is a (reversible) Yang-Baxter map \cite{APT04a,CZ14}. Conversely, the matrix refactorization of the transmission coefficients associated to the solitons provides an alternative method to show that the map \eqref{e:Manakov_p} is a Yang-Baxter map \cite{GV03,V03}\footnote{There is a vast literature on Yang-Baxter maps in relation to (fully) discrete integrable systems and consistency around the cube that would be too long to cover here and also not directly relevant to the present work. We refer the interested reader to the book \cite{HJN16}.}, that is, a solution of the set-theoretical Yang-Baxter equation \cite{D92}.

Like the Manakov system itself,  the discretization  \eqref{e:idMq} is completely integrable in the sense that it can be solved via the IST and it has exact multisoliton solutions \cite{TWU99,AOT99,APT2004,APT04b,APT06}.
As it turns out, \eqref{e:idMq} is even richer, both in terms of the types of soliton solutions it exhibits and, as we will show, in terms of the relative interaction properties. 
%In order to utilize the results derived in the framework of the IST in earlier works \cite{APT2004}, 
As shown in \cite{APT2004}, it is convenient to rewrite the system \eqref{e:idMq} for a $2\times 2$ matrix potential
\begin{equation}
\label{e:Qn}
\@Q_n(\tau)=\begin{pmatrix} 
Q_n^{(1)} & Q_n^{(2)} \\
(-1)^{n+1}Q_n^{(2)\, *} & (-1)^n Q_n^{(1)\, *} 
\end{pmatrix}\,,
\end{equation}
satisfying:
\begin{gather}
\label{e:idMQ}
i\frac{d \@Q_n}{d\tau}=\@Q_{n+1}+ \@Q_{n-1}+\| \@Q_n\|^2 \left(\@Q_{n+1}+\@Q_{n-1}\right)\,,
\end{gather}
where we define the matrix norm as $\|\@Q_n\|^2=|Q_n^{(1)}|^2+|Q_n^{(2)}|^2 $, and let 
\begin{equation}
\@q_n=e^{-2i\tau}h^{-1}(Q_n^{(1)},Q_n^{(2)})^T, \qquad t=\tau/h^2.
\end{equation}

The system \eqref{e:idMQ} admits a variety of discrete vector soliton solutions, depending on the rank
and structure of the norming constant associated with the soliton: fundamental solitons,
fundamental breathers (a superposition of two orthogonally polarized fundamental solitons with the same amplitude and velocity, but opposite carrier frequencies), as well as composite breathers (more general superpositions of fundamental solitons). Fundamental solitons are the discrete analog of \eqref{e:1solitonM}, to which they reduce in the continuous limit. On the other hand, fundamental and composite breathers are purely discrete solutions, and do not have a continuous counterpart in the Manakov system.
One of the goals of this study is to obtain a complete characterization of soliton-breather and breather-breather interactions for the integrable discrete Manakov system. The  formulas for the polarization shifts of discrete fundamental solitons
that are the analog of the well-known formulas \eqref{e:Manakov_p} for
the interaction of vector solitons in the Manakov system were already obtained in \cite{APT2004,APT04b,APT06}.
In this work, we complete the description of the landscape of interactions in the discrete Manakov model by characterizing the interactions between a fundamental soliton and a fundamental breather, between two fundamental breathers, between composite breathers, and between a composite breather and a fundamental soliton or breather.
%In this work, we consider more general interactions of a fundamental soliton and a fundamental breather, and between two fundamental breathers.
Our study reveals that upon interacting with a fundamental breather, a fundamental soliton becomes a fundamental breather and, conversely, that the interaction of two fundamental breathers generically yields two fundamental breathers with polarization shifts, but may also result in a fundamental soliton and a fundamental breather. This type of highly non-trivial interaction was discovered for vector solitons of the complex-coupled short pulse equation (ccSPE) in \cite{CGP23}, and it is now reported for the first time for a discrete integrable system of NLS-type. Explicit formulas for the coefficients that characterize the fundamental breathers and for their polarization vectors are also obtained. 
Furthermore, we show that composite breathers interact trivially both with each other and with a fundamental soliton or breather. All interactions are illustrated with plots.
The results are then interpreted in terms of a refactorization property for the transmission coefficients associated to each soliton, which produces a novel Yang-Baxter map of trigonometric type.

The structure of the paper is as follows. In Section~2, we give a brief overview of the IST for the IDM system as developed in Ref.~\cite{APT2004}, and of its 1-soliton and 1-breather solutions. In Section~3, we derive the explicit expressions of the (matrix) transmission coefficients corresponding to a 1-fundamental soliton, a 1-fundamental breather, and a 1-composite breather solution. In Section~4 we perform long-time asymptotic analysis on exact soliton-breather and breather-breather solutions, and we use it in conjunction with the transmission coefficients for a single soliton/breather solution to obtain the maps for the polarization vectors that describe the solitons and breathers in the forward long-time limit $\tau\to +\infty$ in terms of their values as  $\tau\to -\infty$.  In Section~5 we show, on one hand, how the fundamental breather interaction leads to a Yang-Baxter refactorization property for the transmission coefficients, and, on the other hand, how the map itself can be derived from the refactorization property. Finally, Section~6 is devoted to some concluding remarks, and more technical details are provided in the appendices.

\section{Overview of the IST and soliton/breather solutions}

Below, we give a succinct overview of the IST for the integrable discrete Manakov system \eqref{e:idMQ} as developed in \cite{APT2004}, whose notations we will follow unless specified otherwise. We refer the reader to \cite{APT2004} for further details regarding the results summarized in this section.

Eq.~\eqref{e:idMQ} admits the following Lax pair:
\bse
\label{e:Lax}
\begin{gather}
\label{e:Lax1}
\@v_{n+1}=\begin{pmatrix}
z \@I_2 & \@Q_n \\
\@R_n & z^{-1} \@I_2
\end{pmatrix}\@v_n\,, \\
\label{e:Lax2}
\frac{d}{d\tau}\@v_n=
\begin{pmatrix}
i\@Q_n\@R_{n-1}-\frac{i}{2}\left(z^2+z^{-2}\right)\@I_2 &
-iz\@Q_n+iz^{-1}\@Q_{n-1} \\
iz^{-1}\@R_n-iz\@R_{n-1} & -i\@R_n\@Q_{n-1}+\frac{i}{2}\left(z^2+z^{-2}\right)\@I_2 
\end{pmatrix}\@v_n\,,
\end{gather}
\ese
where 
\begin{equation}
\label{e:Q,R}
\@Q_n(\tau)=\begin{pmatrix} 
Q_n^{(1)} & Q_n^{(2)} \\
(-1)^{n+1}Q_n^{(2)\, *} & (-1)^n Q_n^{(1)\, *} 
\end{pmatrix}, \qquad
\@R_n=(-1)^{n+1} \boldsymbol\sigma_{2} \@Q_n^T \boldsymbol\sigma_{2}\,, \qquad\boldsymbol\sigma_{2}=\begin{pmatrix} 0 & -i \\
i & 0\end{pmatrix},
\end{equation}
$z\in \Complex$ is the spectral parameter, and $\@I_2$ is the $2\times 2$ identity matrix. Note that this corresponds to taking $N=M=2$, $\@R_n=-\@Q_n^\dagger$, and $\@A=\@B=\@I_2$ in \cite{APT2004}; and that the matrix $\@P$ used throughout \cite{APT2004} is simply $i\boldsymbol\sigma_{2}$.

\subsection{Direct problem}
First, one needs to characterize the spectrum
of the scattering problem, namely, Eq.~\eqref{e:Lax1}, and the corresponding eigenfunctions. In the direct and in the inverse problems, $\tau$ is fixed and therefore in the corresponding sections we omit the $\tau$-dependence of the eigenfunctions and scattering data for brevity. Assuming $\@Q_n\to 0$ sufficiently rapidly as $n\to \pm \infty$, one introduces Jost eigenfunctions:
\begin{equation}
\boldsymbol{\Phi}_n(z)=\left(  \boldsymbol\phi_n(z)\ \
\bar{\boldsymbol\phi}_n(z)\right), \qquad \boldsymbol{\Psi}_n(z)=\left(
\bar{\boldsymbol\psi}_n(z)\ \ \boldsymbol\psi_n(z)\right), \label{PhiPsi}
\end{equation}
which  are defined by
\label{eigenf}
\begin{gather}
\boldsymbol{\Phi}_n(z)\sim 
%\@I_4 
\@Z^n\, %e^{i\boldsymbol{\omega\tau\Sigma}_3} 
\qquad n\to -\infty, \qquad
\boldsymbol{\Psi}_n(z)\sim 
%\@I_4
\@Z^n\, %e^{i\omega\tau\boldsymbol{\Sigma}_3} 
\qquad n\to +\infty,
\end{gather}
where
\vspace*{-5mm}
\begin{gather}
\@Z=
\begin{pmatrix} z\, \@I_2 & \@0_2 \\
\@0_2 & z^{-1} \@I_2
    \end{pmatrix},
\end{gather}
with $\@0_2$ being the $2\times 2$ zero matrix.
It is convenient to work with modified eigenfunctions
\label{def_Ms,Ns}
\begin{gather}
\left(  \@M_n(z) \ \bar{\@M}_n(z) \right)
=\boldsymbol{\Phi}_n(z)
%e^{-i\omega(z) \tau \Sigma_3}
\@Z^{-n},
\qquad
\left( \bar{\@N}_n(z)\ \@N_n(z)\right)
=\boldsymbol{\Psi}_n(z)
%e^{-i\Omega(z) \tau \Sigma_3}
\@Z^{-n},
\end{gather}
both approaching the $4\times 4$ identity as $n\to \mp \infty$, respectively.
Let $D^\mp=\left\{z\in \Complex : |z|\lessgtr 1\right\}$ denote the interior ($-$) and the exterior ($+$) of the unit circle $\mathcal{C}=\left\{z\in \Complex: |z|=1\right\}$. As shown in \cite{APT2004}, if the potential $\@Q_n\in \ell^1(\mathbb{Z})$ (i.e., with $\sum_{n=-\infty}^{+\infty}\|\@Q_n\|_a<\infty$ where $\|\@Q_n\|_a$ is any matrix norm of $\@Q_n$), then $\@M_n(z)$, $\@N_n(z)$ defined above are analytic in $D^+$ and continuous for $|z|\ge 1$, and  $\bar{\@M}_n(z)$, $\bar{\@N}_n(z)$ are analytic for $z\in D^-$ and continuous for $|z|\le 1$. Furthermore, the modified eigenfunctions satisfy the following asymptotics (in their respective regions of analyticity):
\bse
\label{e:efs_asym}
\begin{gather}
\@M_n(z)\underset{z\rightarrow\infty}{\sim}
\begin{pmatrix}
\@I_2+\mathcal{O}(z^{-2},\text{even}) \\
-z^{-1} \@Q_{n-1}^\dagger+\mathcal{O}(z^{-3},\text{odd})
\end{pmatrix}, \qquad
\bar{\@M}_n(z)\underset{z\rightarrow0}{\sim}
\begin{pmatrix}
z\, \@Q_{n-1}+\mathcal{O}(z^{3},\text{odd}) \\
\@I_2+\mathcal{O}(z^2,\text{even})
\end{pmatrix}, \\
\@N_n(z)\underset{z\rightarrow\infty}{\sim}
\begin{pmatrix}
-z^{-1} \Delta_n^{-1} \@Q_{n}+\mathcal{O}(z^{-3},\text{odd}) \\
 \Delta_n^{-1}\@I_2+\mathcal{O}(z^{-2},\text{even}) \\
\end{pmatrix}, \qquad
\bar{\@N}_n(z)\underset{z\rightarrow0}{\sim}
\begin{pmatrix}
\Delta_n^{-1}\@I_2+\mathcal{O}(z^2,\text{even})\\
z\,\Delta_n^{-1} \@Q_{n}^\dagger+\mathcal{O}(z^{3},\text{odd})
\end{pmatrix}, 
\end{gather}
\ese
where ``even'' (resp., ``odd'') indicates that the remaining powers are even (resp., ``odd'') powers of $z$, and
\begin{equation}
\label{e:Delta}
\Delta_n=\prod_{k=n}^{+\infty}\left(1+\alpha_k\right)\,, \qquad\alpha_k=|Q_k^{(1)}|^2+|Q_k^{(2)}|^2,
\end{equation}
and we have taken into account that $\@Q_n\@R_n=\@R_n\@Q_n\equiv %\left(|Q_n^{(1)}|^2+|Q_n^{(2)}|^2\right)
\alpha_n\@I_2$ for  $\@Q_n$, $\@R_n$ as in \eqref{e:Q,R}.

The Jost eigenfunctions $\boldsymbol{\Phi}_n(z)$ and $\boldsymbol{\Psi}_n(z)$ are two fundamental matrix solutions of the scattering problem for any $z\in \mathcal{C}$,
%$z\in \Complex$ with $|z|=1$, 
and therefore one can express one in terms of the other as:
\begin{gather}
\label{e:S}
\boldsymbol{\Phi}_n(z)=\boldsymbol{\Psi}_n(z) 
\begin{pmatrix}
\@a(z) & \bar{\@b}(z) \\
\@b(z) & \bar{\@a}(z)
\end{pmatrix}, \qquad
\boldsymbol{\Psi}_n(z)=\boldsymbol{\Phi}_n(z) 
\begin{pmatrix}
\bar{\@c}(z) & \@d(z) \\
\bar{\@d}(z) & \@c(z)
\end{pmatrix}, \qquad z\in \mathcal{C},
\end{gather}
where the $2\times 2$ matrices $\@a(z)$, $\@b(z)$, $\bar{\@a}(z)$, $\bar{\@b}(z)$ are the ``left'' scattering coefficients, and 
the $2\times 2$ matrices $\@c(z)$, $\@d(z)$, $\bar{\@c}(z)$, $\bar{\@d}(z)$ are the ``right'' scattering coefficients.
The scattering coefficients $\@a(z)$, $\@c(z)$ (resp.,  $\bar{\@a}(z)$, $\bar{\@c}(z)$) are analytic for $z\in D^+$ (resp., $z\in D^-$). Moreover, all four diagonal blocks are even functions of $z$ in their respective regions of analyticity, and $\@a(z),\@c(z) \to \@I_2$ as $z\to \infty$. These analytic coefficients are the inverses of the matrix transmission coefficients of the scattering problem.
The off-diagonal scattering coefficients are in general only defined on the unit circle $\mathcal{C}$, where they determine (matrix) reflection coefficients:
\begin{equation}
\boldsymbol{\rho}(z)=\@b(z)\, \@a^{-1}(z), \qquad
\bar{\boldsymbol{\rho}}(z)=\bar{\@b}(z)\, \bar{\@a}^{-1}(z), \qquad |z|=1.
\end{equation}
The symmetries in the Lax pair induce the following symmetries in the scattering coefficients (see \cite{APT2004} for details):
\bse
\begin{gather}
\label{ac_sym}
\bar{\@a}^\dagger(1/z^*)=\@c(z) \prod_{n=-\infty}^{+\infty}(1+\alpha_n), \qquad \@a(z)=\bar{\@c}^\dagger(1/z^*) \prod_{n=-\infty}^{+\infty}(1+\alpha_n),\\
\det \@c(z)=\det \@a(z)\,\prod_{n=-\infty}^{+\infty}(1+\alpha_n)^{-2},\qquad 
\det \@a(z)=\det \bar{\@a}(i/z)=\det \bar{\@c}(i/z)\,\prod_{n=-\infty}^{+\infty}(1+\alpha_n)^{2},
\end{gather}
in the respective regions of analyticity, as well as
\begin{equation}
\bar{\boldsymbol{\rho}}(z)=-\boldsymbol{\rho}^\dagger(1/z^*), \qquad |z|=1.
\end{equation}
\ese
The discrete spectrum consists of the values of $z \in \mathbb{C}\setminus \mathcal{C}$, for which the scattering problem admits eigenfunctions in $\ell^2(\mathbb{Z})$, and, because of the symmetries of the Lax pair, discrete eigenvalues appear in symmetric octets:
\begin{gather}
\label{e:octets}
\mathcal{Z} = \left\{ \pm z_j, \pm \tilde{z}_j:=\pm iz_j^*, \pm \bar{z}_j:=\pm 1/z_j^*, \pm \hat{z}_j:=\pm i/z_j\right\}_{j=1}^J,
\end{gather}
where $\pm z_j, \pm\tilde{z}_j$ are zeros of $\det \@a(z)$ (as well as $\det \@c(z)$) in $D^+$, and 
 $\pm \bar{z}_j, \pm\hat{z}_j$  are zeros of $\det \bar{\@a}(z)$ (as well as $\det \bar{\@c}(z)$) in $D^-$.
The simplest soliton solutions are obtained assuming that the discrete eigenvalues are simple zeros of $\det \@a(z)$. But it is possible for a discrete eigenvalue to be a second-order zero of $\det \@a(z)$ yet still a first-order pole for the meromorphic eigenfunction that appears in the inverse problem. In this respect, both cases should be considered as corresponding to elementary (as opposed to higher order) soliton solutions. 
%$\boldsymbol{\phi}_n(z)\@a^{-1}(z)$. 
Indeed, the following holds.\footnote{Analogous results were established in \cite{POVG18} for the matrix NLS equation and in \cite{GPFT21} for the ccSPE, and similar arguments can be used here.}
\begin{proposition} Let $\left\{\pm z_j, \pm \bar{z}_j,\pm \tilde{z}_j,\pm \hat{z}_j \right\}$ be an octet of discrete eigenvalues as in \eqref{e:octets}.
\begin{enumerate}
\item If $\pm z_j$ are simple zeros of $\det \@a(z)$ in $D^+$, then $\mathrm{rank}\,\@a(\pm z_j)=1$ and 
$\pm z_j$ are simple poles for $\@M_n(z)\@a^{-1}(z)$
(and the same holds for the other symmetric eigenvalues in the octet in the respective regions of analyticity).
\item
If $\pm z_j$ are double zeros of $\det \@a(z)$ in $D^+$ and $\@a(\pm z_j)=\@0_{2}$, then $\pm z_j$ are still simple poles for $\@M_n(z)\@a^{-1}(z)$
(and the same holds for the other symmetric eigenvalues in the octet in the respective regions of analyticity).
\end{enumerate}
Since in both cases the points $\pm z_j,\pm \tilde{z} _j$ (resp., $\pm \bar{z}_j,\pm \hat{z}_j$) are simple poles for the function $\@M \@a^{-1}$ (resp., $\bar{\@M} \bar{\@a}^{-1}$) in $D^{+}$ (resp., $D^{-}$), one can define the corresponding residues as follows:
\begin{subequations}
\label{e:res}
\begin{gather}
\underset{z=\pm z_j}{\mathrm{Res}}\@M_n(z) \@a^{-1}(z) = 
(\pm z_j)^{-2n}\@N_n(\pm z_j)\@C_j,\\
\underset{z=\pm \tilde{z}_j}{\mathrm{Res}}\@M_n(z) \@a^{-1}(z) = 
(\pm i /\bar{z}_j)^{-2n}\@N_n(\pm i/\bar{z}_j)\tilde{\@C}_j,\\
\underset{z=\pm \bar{z}_j}{\mathrm{Res}}\bar{\@M}_n(z) \bar{\@a}^{-1}(z) = (\pm \bar{z}_j)^{2n} \bar{\@N}_n(\pm \bar{z}_j)\bar{\@C}_j,\\
\underset{z=\pm \hat{z}_j}{\mathrm{Res}}\bar{\@M}_n(z) \bar{\@a}^{-1}(z) = (\pm i/z_j)^{2n} \bar{\@N}_n( \pm i/z_j)\tilde{\@C}_n,
\end{gather}
\end{subequations}
where $\@C_j$ is the $2\times2$ norming constant associated to the discrete eigenvalues $\pm z_j$, and
\begin{gather}
\label{e:Csymm}
\bar{\@C}_j = \bar{z}_j^{2} \@C_j^\dagger, \quad \tilde{\@C}_j = -\bar{z}_j^{-2}\boldsymbol\sigma_{2} \bar{\@C}_j^T \boldsymbol\sigma_{2}, \quad \hat{\@C}_j =  z_j^{-2}\boldsymbol\sigma_{2} \@C_j^T \boldsymbol\sigma_{2}\,.
\end{gather}
In the first case, i.e., when $\@a(z),\bar{\@a}(z)$ evaluated at the discrete eigenvalues are rank-1 matrices, the norming constants are rank-1 matrices themselves; in the second case, the norming constants can be either full-rank or rank-one matrices. 
\end{proposition}
As will be explained in Section~2.4, the nature of the soliton associated with a discrete eigenvalue $z_j$ crucially depends on the rank of the associated norming constant $\@C_j$. 

\subsection{Inverse problem} 
The inverse problem aims at reconstructing the potential $\@Q_n$ in terms of the scattering data (i.e., reflection coefficients, discrete eigenvalues, and associated norming constants). Concretely, one first reconstructs the eigenfunctions from the scattering data, and then the potential is recovered from the asymptotics of the eigenfunctions in the spectral parameter $z$. The inverse problem for the eigenfunctions is formulated as a Riemann-Hilbert problem (RHP) in the complex variable $z$, for which a suitable normalization condition as $z\to \infty$ must be provided. Since the large-$z$ asymptotic behavior of the modified eigenfunctions depends on the potential $\@Q_n$ (cf. Eqs.~\eqref{e:efs_asym}), it is convenient to pose the problem for the following renormalized matrix functions:
\bse
\label{e:Nprimes}
\begin{gather}
\@N_n^{\prime}(z)=\begin{pmatrix} \@I_2 & \@0_2 \\ \@0_2 & \Delta_n \@I_2\end{pmatrix}\@N_n(z)
\underset{z\rightarrow\infty}{\sim}\begin{pmatrix} -z^{-1} \Delta_n^{-1} \@Q_n+\mathcal O(z^{-3}) \\
\@I_2+\mathcal O(z^{-2})\end{pmatrix}, \\ 
\bar{\@N}_n^{\prime}(z)=\begin{pmatrix} \@I_2 & \@0_2 \\ 
\@0_2 & \Delta_n \@I_2\end{pmatrix}\bar{\@N}_n(z)
\underset{z\rightarrow0}{\sim}\begin{pmatrix} \Delta_n^{-1}\@I_2 +\mathcal O(z^2)\\
z\,\@Q_n^\dagger+\mathcal O(z^{3})\end{pmatrix}, \\
\boldsymbol{\mu}_n^{\prime}(z)=\begin{pmatrix} \@I_2 & \@0_2 \\ \@0_2 & \Delta_n \@I_2
\end{pmatrix}\@M_n(z)\@a^{-1}(z)\underset{z\rightarrow\infty}{\sim}\begin{pmatrix}\@I_2+\mathcal O(z^{-2}) \\ -z^{-1}\Delta_n \@Q_{n-1}^\dagger+\mathcal O(z^{-3})
\end{pmatrix}, \\
\bar{\boldsymbol{\mu}}_n^{\prime}(z)=\begin{pmatrix} \@I_2 & \@0_2 \\ \@0_2 & \Delta_n \@I_2\end{pmatrix}\bar{\@M}_n(z)\bar{\@a}^{-1}(z)
\underset{z\rightarrow0}{\sim}\begin{pmatrix} z\@Q_{n-1}+\mathcal O(z^3) \\
\Delta_n \@I_2+\mathcal O(z^2)\end{pmatrix}.
\end{gather}
\ese
These eigenfunctions are meromorphic functions of $z$ with poles (assumed simple, cf. Proposition~1) at the discrete eigenvalues, and 
the $4\times 4$ matrix function $\@m_n(z)=\@m_n^\pm (z)$ for $z\in D^\pm$:
$$
\@m_n^+(z)=\left( \boldsymbol{\mu}_n^\prime(z)\  \@N_n^\prime(z)\right), \qquad 
\@m_n^-(z)=\left(\bar{\@N}_n^\prime(z)\ \bar{\boldsymbol{\mu}}_n^\prime(z)\right),
$$
satisfies the following RHP across the circle $|z|=1$:
\bse
\begin{gather}
\label{e:RHP}
\@m_n^+(z)=\@m_n^-(z)\left(\@I_4+\@V_n(z)\right) , \qquad
\@V_n(z)=\begin{pmatrix} \boldsymbol{\rho}^\dagger(1/z^*)
\boldsymbol{\rho}(z) & z^{2n}\boldsymbol{\rho}^\dagger(1/z^*) \\
z^{-2n} \boldsymbol{\rho}(z) & \@0_2\end{pmatrix}, \\
\@m_n^+(z)\to \@I_4 \qquad \text{as } z\to \infty,
\end{gather}
\ese
with simple poles at the discrete eigenvalues whose residues are determined by the norming constants according to \eqref{e:res}. In turn, the potential $\@Q_n$ is reconstructed by the asymptotics of the renormalized eigenfunctions, namely:
\begin{equation}
\@Q_{n-1}=\lim_{z\to 0}z^{-1}\, \bar{\boldsymbol{\mu}}_n^{\prime\, \textrm{(up)}}(z),
\qquad \Delta_n \@I_2=\lim_{z\to 0} \bar{\boldsymbol{\mu}}_n^{\prime\, \textrm{(dn)}}(z), 
\end{equation}
where, here and in the following, the superscripts $^{\textrm{(up)}}$ and $^{\textrm{(dn)}}$ denote the upper/lower $2\times 2$ block of the $4\times 2$ matrix eigenfunctions.
Note that as a consequence of the symmetries of the scattering data (and consistently with the expansions \eqref{e:efs_asym}), one has:
\bse
\label{e:symmefs}
\begin{gather}
\bar{\@N}_n^{\prime\, \textrm{(up)}}(-z)=\bar{\@N}_n^{\prime\, \textrm{(up)}}(z), \qquad
\bar{\@N}_n^{\prime\, \textrm{(dn)}}(-z)=-\bar{\@N}_n^{\prime\, \textrm{(dn)}}(z), \\
\@N_n^{\prime\, \textrm{(up)}}(-z)=-\@N_n^{\prime\, \textrm{(up)}}(z), \qquad
\@N_n^{\prime\, \textrm{(dn)}}(-z)=\@N_n^{\prime\, \textrm{(dn)}}(z).
\end{gather}
\ese
In the pure soliton case (i.e., for reflectionless potentials for which $\boldsymbol{\rho}(z)\equiv 0$), the jump in \eqref{e:RHP} becomes trivial, and, taking into account the symmetries \eqref{e:symmefs}, the solution for the RHP for the eigenfunctions is given by:
\bse
\label{Nbar_RHP}
\begin{gather}
\bar{\@N}_n^{\prime\, \textrm{(up)}}(z)=\@I_2 
 +2\sum_{j=1}^J \frac{z_j^{-2n+1}}{z^2-z_j^2} 
 \@N_n^{\prime\, \textrm{(up)}}(z_j)\@C_j+2i\sum_{j=1}^J\frac{(-1)^n\bar{z}_j^{2n-1}}{z^2+\bar{z}_j^{-2}}\@N_n^{\prime\, \textrm{(up)}}(i/\bar{z}_j)\tilde{\@C}_j, \\
\bar{\@N}_n^{\prime\, \textrm{(dn)}}(z)= 
 2\sum_{j=1}^J \frac{z_j^{-2n}z}{z^2-z_j^2} 
\@N_n^{\prime\, \textrm{(dn)}}(z_j)\@C_j+2\sum_{j=1}^J\frac{(-1)^n\bar{z}_j^{2n}z}{z^2+\bar{z}_j^{-2}}\@N_n^{\prime\, \textrm{(dn)}}(i/\bar{z}_j)\tilde{\@C}_j, \\
\@N_n^{\prime\, \textrm{(up)}}(z)=
 2\sum_{j=1}^J \frac{\bar{z}_j^{2n}z}{z^2-\bar{z}_j^2} 
 \bar{\@N}_n^{\prime\, \textrm{(up)}}(\bar{z}_j)\bar{\@C}_j+2\sum_{j=1}^J\frac{(-1)^n z_j^{-2n}z}{z^2+z_j^{-2}}\bar{\@N}_n^{\prime\, \textrm{(up)}}(i/z_j)\hat{\@C}_j,  \\
\@N_n^{\prime\, \textrm{(dn)}}(z)=
\@I_2+ 2\sum_{j=1}^J \frac{\bar{z}_j^{2n+1}}{z^2-\bar{z}_j^2} 
 \bar{\@N}_n^{\prime\, \textrm{(dn)}}(\bar{z}_j)\bar{\@C}_j+2i\sum_{j=1}^J\frac{(-1)^n z_j^{-2n-1}}{z^2+z_j^{-2}}\bar{\@N}_n^{\prime\, \textrm{(dn)}}(i/z_j)\hat{\@C}_j, 
\end{gather}
\ese
which yields a closed linear system for the eigenfunctions evaluated at the discrete eigenvalues, i.e.:
\bse
\label{up_system}
\begin{gather}
\label{up_system_1}
\bar{\@N}_n^{\prime\, \textrm{(up)}}(\bar{z}_j)
=\@I_2+2\sum_{k=1}^J\frac{z_k^{-2n+1}}{\bar{z}_j^2-z_k^2}
\@N_n^{\prime\, \textrm{(up)}}(z_k)\@C_k
+2i\sum_{k=1}^J\frac{(-1)^n\bar{z}_k^{2n-1}}{\bar{z}_j^2+\bar{z}_k^{-2}}
\@N_n^{\prime\, \textrm{(up)}}(i/\bar{z}_k)\tilde{\@C}_k,\\
\label{up_system_2}
\bar{\@N}_n^{\prime\, \textrm{(up)}}(i/z_j)
=\@I_2-2\sum_{k=1}^J\frac{z_k^{-2n+1}}{z_j^{-2}+z_k^2}
\@N_n^{\prime\, \textrm{(up)}}(z_k)\@C_k
-2i\sum_{k=1}^J\frac{(-1)^n\bar{z}_k^{2n-1}}{z_j^{-2}-\bar{z}_k^{-2}}
\@N_n^{\prime\, \textrm{(up)}}(i/\bar{z}_k)\tilde{\@C}_k, \\
\label{up_system_3}
\@N_n^{\prime\, \textrm{(up)}}(z_j)
=2\sum_{k=1}^J\frac{\bar{z}_k^{2n}z_j}{z_j^2-\bar{z}_k^2}
\bar{\@N}_n^{\prime\, \textrm{(up)}}(\bar{z}_k)\bar{\@C}_k
+2\sum_{k=1}^J\frac{(-1)^n z_k^{-2n}z_j}{z_j^2+z_k^{-2}}
\bar{\@N}_n^{\prime\, \textrm{(up)}}(i/z_k)\hat{\@C}_k,\\
\label{up_system_4}
\@N_n^{\prime\, \textrm{(up)}}(i/\bar{z}_j)
=-2i\sum_{k=1}^J\frac{\bar{z}_k^{2n}\bar{z}_j^{-1}}{\bar{z}_j^{-2}+\bar{z}_k^2}
\bar{\@N}_n^{\prime\, \textrm{(up)}}(\bar{z}_k)\bar{\@C}_k
-2i\sum_{k=1}^J\frac{(-1)^n z_k^{-2n}\bar{z}_j^{-1}}{\bar{z}_j^{-2}-z_k^{-2}}
\bar{\@N}_n^{\prime\, \textrm{(up)}}(i/z_k)\hat{\@C}_k,
\end{gather}
\ese
$j=1,\cdots, J$, and
\begin{equation}
\label{potential}
\@Q_{n-1}=-2\sum_{j=1}^J\bar{z}_j^{2(n-1)}
\bar{\@N}_n^{\prime\, \textrm{(up)}}(\bar{z}_j)\bar{\@C}_j
-2\sum_{j=1}^J(-1)^{n-1}z_j^{-2(n-1)}\bar{\@N}_n^{\prime\, \textrm{(up)}}(i/z_j)\hat{\@C}_j.
\end{equation}
For future reference, we also give the linear system for the lower blocks of the renormalized eigenfunctions:
\bse
\label{dn_system}
\begin{gather}
\bar{\@N}_n^{\prime\, \textrm{(dn)}}(\bar{z}_j)
=2\sum_{k=1}^J\frac{z_k^{-2n}\bar{z}_j}{\bar{z}_j^2-z_k^2}
\@N_n^{\prime\, \textrm{(dn)}}(z_k)\@C_k
+2\sum_{k=1}^J\frac{(-1)^n\bar{z}_k^{2n}\bar{z}_j}{\bar{z}_j^2+\bar{z}_k^{-2}}
\@N_n^{\prime\, \textrm{(dn)}}(i/\bar{z}_k)\tilde{\@C}_k,\\
\bar{\@N}_n^{\prime\, \textrm{(dn)}}(i/z_j)
=-2i\sum_{k=1}^J\frac{z_k^{-2n}z_j^{-1}}{z_j^{-2}+z_k^2}
\@N_n^{\prime\, \textrm{(dn)}}(z_k)\@C_k
-2i\sum_{k=1}^J\frac{(-1)^n\bar{z}_k^{2n}z_j^{-1}}{z_j^{-2}-\bar{z}_k^{-2}}
\@N_n^{\prime\, \textrm{(dn)}}(i/\bar{z}_k)\tilde{\@C}_k, \\
\@N_n^{\prime\, \textrm{(dn)}}(z_j)
=\@I_2+2\sum_{k=1}^J\frac{\bar{z}_k^{2n+1}}{z_j^2-\bar{z}_k^2}
\bar{\@N}_n^{\prime\, \textrm{(dn)}}(\bar{z}_k)\bar{\@C}_k
+2i\sum_{k=1}^J\frac{(-1)^n z_k^{-2n-1}}{z_j^2+z_k^{-2}}
\bar{\@N}_n^{\prime\, \textrm{(dn)}}(i/z_k)\hat{\@C}_k,\\
\@N_n^{\prime\, \textrm{(dn)}}(i/\bar{z}_j)
=\@I_2-2\sum_{k=1}^J\frac{\bar{z}_k^{2n+1}}{\bar{z}_j^{-2}+\bar{z}_k^2}
\bar{\@N}_n^{\prime\, \textrm{(dn)}}(\bar{z}_k)\bar{\@C}_k
-2i\sum_{k=1}^J\frac{(-1)^n z_k^{-2n-1}}{\bar{z}_j^{-2}-z_k^{-2}}
\bar{\@N}_n^{\prime\, \textrm{(dn)}}(i/z_k)\hat{\@C}_k,
\end{gather}
\ese
with
\begin{equation}
\label{e:Deltan}
\Delta_n\@I_2=\@I_2-2\sum_{j=1}^J\bar{z}_j^{2n-1}
\bar{\@N}_n^{\prime\, \textrm{(dn)}}(\bar{z}_j)\bar{\@C}_j
+2i\sum_{j=1}^J(-1)^nz_j^{-2n+1}
\bar{\@N}_n^{\prime\, \textrm{(dn)}}(i/z_j)\hat{\@C}_j.
\end{equation}
\subsection{Time evolution} 
The second operator in the Lax pair \eqref{e:Lax2} determines the time-dependence of eigenfunctions and scattering data. Specifically, for the latter the following holds.
\begin{enumerate}
\item The matrix transmission coefficients are constants of the motion:
\begin{equation}
 \@a(z,\tau)=\@a(z,0), \qquad \bar{\@a}(z,\tau)=\bar{\@a}(z,0),   
\end{equation}
so in the following we will continue to denote them simply as $\@a(z)$, $\bar{\@a}(z)$. The same of course holds for $\@c(z)$ and $\bar{\@c}(z)$.
In particular, note that the above equations imply that the discrete eigenvalues are all time independent.
\item The time-evolution of the matrix reflection coefficient and of the norming constants are given by
\begin{equation}
\boldsymbol{\rho}(z,\tau)=e^{i(z^2+z^{-2})\tau}\boldsymbol{\rho}(z,0),
\qquad \@C_j(\tau)=\@C_je^{i(z_j^2+z_j^{-2})\tau}, \quad j=1,\cdots,J,
\end{equation}
(note $\@C_j=\@C_j(0)$) and \eqref{e:Csymm} provide the corresponding expressions for the other norming constants in each octet.
\end{enumerate}
%In the following, we will use  $\@C_j$ to denote %$\@C_j(0)$, %and explicitly indicate the time %dependence, 
%and the same for the other norming constants in each %octet.
\subsection{Explicit 1-soliton and 1-breather solutions}
The simplest soliton solutions are obtained for 1 octet of discrete eigenvalues, i.e., by taking $J=1$, and the nature of the solution crucially depends on the rank of the associated norming constants.
When the norming constant $\@C_1$ is rank-1, and one of its columns is identically zero, say
\begin{equation}
\@C_1=\left( \boldsymbol{\gamma}_1 \ \@0\right), \quad \boldsymbol{\gamma}_1\in \Complex^2,   
\end{equation}
the corresponding solution is a \textbf{fundamental soliton}:
\bse
\begin{equation}
\label{e:1FS}
%\@Q_n(\tau)
\begin{pmatrix}
Q_n^{(1)}(\tau) \\  Q_n^{(2)}(\tau)
\end{pmatrix}= -\sinh2a_{1}\sech(\zeta_{1}-d_{1})e^{2ib_{1}(n+1)-2i\omega_{1}\tau}\frac{\boldsymbol\gamma_{1}^{*}}{\|\boldsymbol\gamma_{1}\|}\,,
\end{equation}
where $z_{1}=\exp(a_{1}+ib_{1})$ (with $a_1>0$ since $z_1\in D^+$), and
\begin{eqnarray}
    &\zeta_{1}=2a_{1}\big((n+1)-v_{1}\tau\big)\,,\;&d_{1}=\log\frac{\|\boldsymbol\gamma_{1}\|}{\sinh2a_{1}}\,,\\
    &\omega_{1}=\cosh2a_{1}\cos2b_{1}\,,\;\;&v_{1}=-\frac{1}{a_{1}}\sinh2a_{1}\sin2b_{1}\,.
\label{e:soliton_params}
\end{eqnarray}
\ese
When the norming constant $\@C_1$ is rank-1 with the two columns proportional to each other, say

\begin{equation}
\label{e:Cfb}
\@C_1= \boldsymbol{\gamma}_1\boldsymbol{\delta}_1^\dagger\equiv\left( \mu_1 \boldsymbol{\gamma}_1, \ \kappa_1  \boldsymbol{\gamma}_1\right), \quad \boldsymbol{\gamma}_1\in \Complex^2, \quad \boldsymbol{\delta}_1=\begin{pmatrix}\mu^{*}_1 \\ \kappa^{*}_1\end{pmatrix}\in \Complex^2,  
\end{equation}
the corresponding solution is a \textbf{fundamental breather}:
\bse
\label{fb}
\begin{equation}
 %\@Q_n(\tau)
 \begin{pmatrix}
  Q_n^{(1)}(\tau) \\ Q_n^{(2)}(\tau)    
 \end{pmatrix}
 =-\sinh2a_{1}\sech(\zeta_{1}-d_{1})\left[\frac{\mu_{1}^{*}}{\|\boldsymbol\delta_{1}\|}\frac{\boldsymbol\gamma_{1}^{*}}{{\|\boldsymbol\gamma_{1}\|}}e^{2ib_{1}(n+1)-2i\omega_{1}\tau}+(-1)^{n+1}\frac{\kappa_{1}}{{\|\boldsymbol\delta_{1}\|}}\frac{\boldsymbol\gamma_{1}^{\perp}}{{{\|\boldsymbol\gamma_{1}\|}}}e^{-2ib_{1}(n+1)+2i\omega_{1}\tau}\right]\,,
\end{equation}
where
\vspace*{-2mm}
\begin{equation}
\boldsymbol\gamma_{1}^{\perp}=-i\boldsymbol\sigma_{2}\boldsymbol\gamma_{1}\,, \qquad d_{1}=\log\frac{\|\boldsymbol\gamma_{1}\|\|\boldsymbol\delta_{1}\|}{\sinh2a_{1}}\,,
\end{equation}
\ese
and the rest of the parameters are the same as in the fundamental soliton case. The above expression shows that a fundamental breather is a superposition of two orthogonally polarized fundamental solitons, with the same amplitude and velocity, but opposite carrier frequencies. We want to stress that the vector $\boldsymbol{\delta}_1$ introduced here is different from the one used in \cite{APT2004} (where $\boldsymbol{\delta}_1$ was used to denote the second column of the norming constant), and instead follows the notation introduced in \cite{CGP23} for the ccSPE.
Eq.~\eqref{fb} shows that the fundamental
breather reduces to a fundamental soliton by setting either $\kappa_1=1$ and $\mu_1=0$, or
$\kappa_1=0$ and $\mu_1=1$. Of course, one of the two constants $\kappa_1$ and $\mu_1$ can always be scaled out (as in the solutions presented in \cite{APT2004}, where $\mu_1=1$ and $\kappa_1=\kappa$). However, for the purpose of
investigating soliton interactions, it is convenient to keep both constants in. It is worth mentioning that if $\boldsymbol{\gamma}_1$ has only one non-zero component, then the opposite carriers are split in different components, and hence the oscillations are suppressed. While with a single breather it is always possible to reduce it to such a case by an appropriate axes rotation and exploit the unitary invariance of the system \eqref{e:idMQ}, this is not possible, in general, with more than one breather.

\begin{figure}[ht!]
\centering
    \includegraphics[width=.8\textwidth]{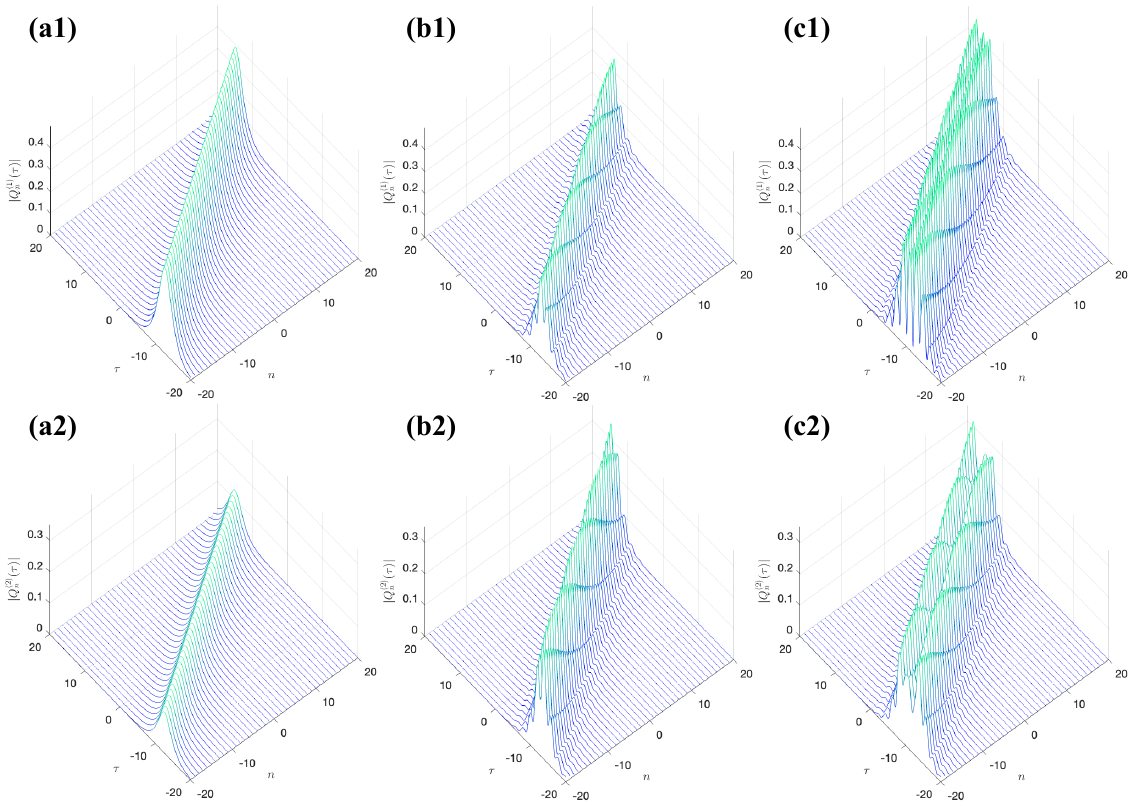}
    \caption{Single soliton solutions corresponding to the same discrete eigenvalue $z_{1}=\exp(0.2-i\pi/8)$, with $|Q_n^{(1)}(\tau)|$ in the top panels, and $|Q_n^{(2)}(\tau)|$ in the bottom panels. \textbf{(a)} Fundamental soliton with $\@C_{1}=\begin{pmatrix}0.3&0\\0.1&0\end{pmatrix}$. \textbf{(b)} Fundamental breather with $\@C_{1}=\begin{pmatrix}0.3&0.3\\0.1&0.1\end{pmatrix}$. \textbf{(c)} Composite breather with $\@C_{1}=\begin{pmatrix}0.3&0.3\\0.1&0.2\end{pmatrix}$.}
    \label{f:1soliton}
\end{figure}

Finally, when the norming constant  $\@C_1=(\boldsymbol{\gamma}_1\ \boldsymbol{\varepsilon}_1)$ is invertible (i.e., its columns $\boldsymbol{\gamma}_1, \boldsymbol{\varepsilon}_1$ are linearly independent), the corresponding solution is a \textbf{composite breather}:
\bse
\label{comp_bre}
\begin{equation}
%\@Q_n(\tau)
 \begin{pmatrix}
  Q_n^{(1)}(\tau) \\ Q_n^{(2)}(\tau)    
 \end{pmatrix}
=\frac{1}{1+g_{n+1}(\tau)}
\Big[\Theta_{n+1}(\tau)
e^{-2i(\omega_1+ia_1v_1)\tau}
%{\@C}_{1}^{\dagger}
\boldsymbol{\gamma}_1^*
+(-1)^{n+1}\Theta_{n+1}^{*}(\tau)e^{2i(\omega_1-ia_1v_1)\tau}
%\@P{\@C}_{1}^{T}\@P
\boldsymbol{\varepsilon}_1^\perp
\Big]\,,
\end{equation}
where %$\omega_{1}=z_{1}^{2}+z_{1}^{-2}$, $\bar\omega_{1}=\bar{z}_{1}^{2}+\bar{z}_{1}^{-2}$ and
\begin{eqnarray}
\label{g_composite}
    &\displaystyle g_{n}(\tau)=\frac{e^{-4a_{1}n+{4a_1v_1\tau}}}{\sinh^{2}2a_{1}}\|\@C_{1}\|^{2}+\Gamma_{n}(\tau)+\Gamma_{n}^{*}(\tau)+s^{4}|\Gamma_{n}(\tau)|^{2}\,, \qquad s=\frac{\cos2b_{1}}{\sinh2a_{1}}\,,\\
    &\displaystyle \Theta_{n}(\tau)=-2\bar{z}_{1}^{2n}\big[1-s^{2}\Gamma_{n}(\tau)\big]\,,\;\;\;\Gamma_{n}(\tau)=(-1)^{n}\frac{1}{(\omega_1-ia_1v_1)^{2}}z_{1}^{-4n}e^{4i(\omega_1-ia_1v_1)\tau}\det\@C_{1}\,.
\end{eqnarray}
\ese
The above formula for a generic composite breather is novel; only a special case, corresponding to $\@C_1=\left(\boldsymbol{\gamma}_1\,, 
i\eta \boldsymbol\sigma_{2} \boldsymbol{\gamma}_1^*\right)$ with $\eta\in \Complex$, was given in \cite{APT04b}. We also mention that in \cite{O09} soliton solutions for the discrete coupled nonlinear Schr\"odinger equations (which are essentially the IDM system considered here) are derived using Hirota's bilinear formalism. However, only fundamental soliton solutions can be obtained from the Pfaffian solutions of the bilinear equations that are reported in \cite{O09}.
Some plots of 1-soliton and 1-breather solutions are given in Fig.~\ref{f:1soliton}.

As we will show in Section~4, the key difference between rank-1 and rank-2 solitons is that the former interact nontrivially with each other and with rank-2 solitons. Conversely, when rank-2 solitons interact with each other or with rank-1 solitons, the only effect of the interaction is an overall phase/position shift, while the polarization vectors are left unchanged.

\section{Transmission coefficients of soliton and breather solutions}

According to their definition in Section 2, the (inverse) transmission coefficients $\@c(z),\,\bar{\@c}(z)$ corresponding to single soliton solutions can be determined from limits of the ``right" modified eigenfunctions as follows:
\begin{eqnarray}
\label{e:lim_c}
    \bar{\@c}(z)=\lim_{n\rightarrow-\infty}\bar{\@N}_{n}^{(\text{up})}(z)\,, \qquad \@c(z)=\lim_{n\rightarrow-\infty}\@N_{n}^{(\text{dn})}(z)\,.
\end{eqnarray}
For instance, from the second of \eqref{e:S} it follows that for $z\in \mathcal{C}$:
$$
\bar{\@N}_{n}^{(\text{up})}(z)=\@M_{n}^{(\text{up})}(z)\bar{\@c}(z)+z^{-2n}\bar{\@M}_{n}^{(\text{up})}(z)\@d(z).
$$
Taking the limit as $n\to -\infty$ in the above equation, and considering that $\@M_{n}^{(\text{up})}(z)\to \@I_2$ and
$\bar{\@M}_{n}^{(\text{up})}(z)\to \@0_2$ one obtains the first of \eqref{e:lim_c}, which can then be extended into $D^-$ by analytic continuation. A similar approach works for $\@c(z)$.
The coefficients $\@a(z),\,\bar{\@a}(z)$ can be found via similar limits of the ``left" eigenfunctions, or directly through the symmetries in \eqref{ac_sym}.

To compute the relevant eigenfunction needed to determine $\bar{\@c}(z)$, one can first solve the linear system \eqref{up_system_1}--\eqref{up_system_4} with $J=1$, and substitute the result into \eqref{Nbar_RHP} to determine $\bar{\@N}_{n}^{\prime\, (\text{up})}(z)$. Eq.~\eqref{e:Deltan} allows one to obtain $\Delta_n$ in terms of the solution of the linear system \eqref{dn_system}, which can then be used in conjunction with \eqref{e:Nprimes} to compute the limit required in \eqref{e:lim_c}.
We provide below the results of these calculations for a fundamental soliton, fundamental breather and composite breather. Further details are given in Appendix~A. 

In the case of a composite breather, where the norming constant $\@C_{1}$ is invertible, the transmission coefficients simply turn out to be multiples of the identity; namely,
\bse
\label{e:CB_transm}
\begin{eqnarray}
    &\bar{\@c}_{\text{CB}}(z)=\displaystyle\frac{(z^{2}-\bar{z}_{1}^{2})(z^{2}+z_{1}^{-2})}{(z^{2}-z_{1}^{2})(z^{2}+\bar{z}_{1}^{-2})}\@I_{2}\,,\;\;\;&
    \@c_{\text{CB}}(z)=\frac{(z^{-2}-{z}_{1}^{-2})(z^{-2}+\bar{z}_{1}^{2})}{(z^{-2}-\bar{z}_{1}^{-2})(z^{-2}+z_{1}^{2})}\@I_{2}\,,\\
    &\@a_{\text{CB}}(z)=\displaystyle\frac{(z^{2}-{z}_{1}^{2})(z^{-2}+\bar{z}_{1}^{2})}{(z^{2}-\bar{z}_{1}^{2})(z^{-2}+z_{1}^{2})}\@I_{2}\,,\;\;\;&
    \bar{\@a}_{\text{CB}}(z)=\frac{(z^{-2}-\bar{z}_{1}^{-2})(z^{2}+{z}_{1}^{-2})}{(z^{-2}-{z}_{1}^{-2})(z^{2}+\bar z_{1}^{-2})}\@I_{2}\,.
\end{eqnarray}
\ese
%As we will see, this implies that composite breathers interact trivially with any other type of soliton/breather.
Note that the above expressions correspond to the second case in Proposition~1, i.e., each of the discrete eigenvalues in the octet is a double zero of the determinant of one of the coefficients, all of which have rank 0 (i.e., they vanish) when evaluated at the eigenvalues in the pertinent region of analyticity.

On the other hand, the transmission coefficients are non-trivial in the rank-1 case. The primary norming constant associated with a fundamental breather can be written as
$\@C_{1}=\boldsymbol\gamma_{1}\boldsymbol\delta_{1}^{\dagger}$; and with the symmetries \eqref{e:Csymm} in mind, one has $\hat{\@C}_{1}\@C_{1}=\bar{\@C}_{1}\tilde{\@C}_{1}=\@0_2$. As such, many terms in the linear system vanish, including those that contribute to the transmission coefficient limits in the rank-2 case. The transmission coefficients for the fundamental breather, derived in Appendix A, are as follows:
\bse
\label{e:FB_transm}
\begin{eqnarray}
\label{cbar}
    \bar{\@c}_{\text{FB}}(z)&=&\frac{z^{2}+z_{1}^{-2}}{z^{2}+\bar{z}_{1}^{-2}}\bigg[\@I_{2}+\frac{(\bar{z}_{1}^{2}+z_{1}^{-2})(\bar{z}_{1}^{-2}-z_{1}^{-2})}{(z_{1}^{-2}-z^{-2})(z^{2}+z_{1}^{-2})}\@v^{*}_{1}\@v_{1}^{T}\bigg]\,,\\
    \label{c}
    \@c_{\text{FB}}(z)&=&\frac{\bar{z}_{1}^{2}+z^{-2}}{z_{1}^{2}+z^{-2}}\bigg[\@I_{2}+\frac{(\bar{z}_{1}^{2}-z_{1}^{2})(\bar{z}_{1}^{2}+z_{1}^{-2})}{(z^{2}-\bar{z}_{1}^{2})(\bar{z}_{1}^{2}+z^{-2})}\@u^{*}_{1}\@u_{1}^{T}\bigg]\,,\\
    \label{a}
    {\@a}_{\text{FB}}(z)&=&\frac{z^{2}+\bar z_{1}^{-2}}{z^{2}+{z}_{1}^{-2}}\bigg[\@I_{2}+\frac{(\bar{z}_{1}^{2}+z_{1}^{-2})({z}_{1}^{2}-\bar z_{1}^{2})}{(\bar z_{1}^{2}-z^{2})(z^{-2}+\bar z_{1}^{2})}\@v^{*}_{1}\@v_{1}^{T}\bigg]\,,\\
    \label{abar}
    \bar{\@a}_{\text{FB}}(z)&=&\frac{{z}_{1}^{2}+z^{-2}}{\bar z_{1}^{2}+z^{-2}}\bigg[\@I_{2}+\frac{({z}_{1}^{-2}-\bar z_{1}^{-2})(\bar{z}_{1}^{2}+z_{1}^{-2})}{(z^{-2}-{z}_{1}^{-2})({z}_{1}^{-2}+z^{2})}\@u^{*}_{1}\@u_{1}^{T}\bigg]\,,
\end{eqnarray}
\ese
where $\@u_{1}=\boldsymbol\gamma_{1}^{*}/\|\boldsymbol\gamma_{1}\|$ and $\@v_{1}=\boldsymbol\delta_{1}^{*}/\|\boldsymbol\delta_{1}\|$ are normalized polarization vectors. Note that when $\@v_{1}=(1,0)^{T}$, $\bar{\@c}(z)$ in \eqref{cbar} and $\@a(z)$ in \eqref{a} reduce to diagonal matrices 
%(which coincide with the corresponding ones in the composite breather case), 
%^I don't think they do, they are diagonal but not multiples of the identity.
while $\@c(z)$ in \eqref{c} and $\bar{\@a}(z)$ in \eqref{abar} remain nontrivial, in agreement with the result obtained in \cite{APT2004} for the fundamental soliton case. The expressions of the transmission coefficients for fundamental and composite breathers were not determined before.

As we will see in Section~4, the nontrivial structure of their transmission coefficients (and their dependence on the norming constants, in addition to the discrete eigenvalues) is responsible for the nontrivial interaction properties of rank-1 solitons, i.e., fundamental solitons and fundamental breathers. 

%{\color{red}Do we want to say anything about interactions of composite breathers? Without DT, it might be too hard to say much, even though the transmission coefficients suggest trivial interactions... Maybe we can put it in the concluding remarks, as something for future investigation.}

%{\color{blue} If we say something about Manakov's method which should lead to a refactorization of the scattering coefficients, then we can treat fundamental and composition breathers on the same footing I think.}

\section{Soliton-breather and breather-breather solutions and their long-time asymptotics}

\subsection{Generic multi-soliton solution}
The explicit expression of a multi-soliton solution can be obtained from the linear system \eqref{up_system}. Specifically, upon substitution of \eqref{up_system_3} and \eqref{up_system_4} into \eqref{up_system_1} and \eqref{up_system_2}, one obtains:
\bse
\label{single_equ}
\begin{equation}
\boldsymbol{\chi}_j=\@I_{2}+\sum_{\ell=1}^{2J}\boldsymbol{\chi}_\ell \@A_{j\ell},\quad j =1,\dots,2J,
\end{equation}
where
\footnotesize
\begin{equation*}
    \@A_{j\ell}=\begin{cases}
    \displaystyle
    4\sum_{k=1}^J\left[\frac{z_k^{-2n+2}\bar{z}_\ell^{2n}}{(\bar{z}_j^2-z_k^2)(z_k^2-\bar{z}_\ell^2)}\bar{\@C}_\ell \@C_k+\frac{(-1)^n\bar{z}_k^{2n-2}\bar{z}_s^{2n}}{(\bar{z}_j^2+\bar{z}_k^{-2})(\bar{z}_k^{-2}+\bar{z}_\ell^2)}\bar{\@C}_s\tilde{\@C}_k\right],\, \ell=1,\dots,J,j=1,\dots,J,\\
    \displaystyle
        4\sum_{k=1}^J\left[\frac{(-1)^nz_k^{-2n+2}z_{\ell-J}^{-2n}}{(\bar{z}_j^2-z_k^2)(z_k^2+z_{\ell-J}^{-2})}\hat{\@C}_{\ell-J}\@C_k+\frac{\bar{z}_k^{2n-2}z_{\ell-J}^{-2n}}
        {(\bar{z}_j^2+\bar{z}_k^{-2})(\bar{z}_k^{-2}-z_{\ell-J}^{-2})}\hat{\@C}_{\ell-J}\tilde{\@C}_k\right],\, \ell=J+1,\dots,2J,j=1,\dots,J,\\
    \displaystyle
-4\sum_{k=1}^J\left[\frac{z_k^{-2n+2}\bar{z}_\ell^{2n}}{(z_{j-J}^{-2}+z_k^2)(z_k^2-\bar{z}_\ell^2)}\bar{\@C}_\ell \@C_k-\frac{(-1)^n\bar{z}_k^{2n-2}\bar{z}_s^{2n}}
    {(\bar{z}_k^{-2}-z_{j-J}^{-2})(\bar{z}_k^{-2}+\bar{z}_\ell^2)}\bar{\@C}_\ell\tilde{\@C}_k\right],\, \ell=1,\dots,J,j=J+1,\dots,2J,\\
    \displaystyle        -4\sum_{k=1}^J\left[\frac{(-1)^nz_k^{-2n+2}z_{\ell-J}^{-2n}}{(z_{j-J}^{-2}+z_k^2)(z_k^2+z_{\ell-J}^{-2})}\hat{\@C}_{\ell-J}\@C_k-\frac{\bar{z}_k^{2n-2}z_{\ell-J}^{-2n}}
        {(\bar{z}_k^{-2}-z_{j-J}^{-2})(\bar{z}_k^{-2}-z_{\ell-J}^{-2})}\hat{\@C}_{\ell-J}\tilde{\@C}_k\right],\,\ell=J+1,\dots,2J,j=J+1,\dots,2J,
    \end{cases}
\end{equation*} \normalsize
 and
\begin{equation}
\boldsymbol{\chi}\equiv (\boldsymbol{\chi}_1,\boldsymbol{\chi}_2,\dots,\boldsymbol{\chi}_{2J})=\left(\bar{\@N}^{\prime\, \textrm{(up)}}_n(\bar{z}_1),\dots,\bar{\@N}^{\prime \, \textrm{(up)}}_n(\bar{z}_J),\bar{\@N}^{\prime \, \textrm{(up)}}_n(i/z_1),\dots,\bar{\@N}^{\prime \, \textrm{(up)}}_n(i/z_J) \right).
\end{equation}
\ese
For convenience, we have dropped the explicit dependence on the lattice variable $n$. Thus, the set of equations \eqref{single_equ} forms a system of linear equations
\begin{equation}\label{main_system}
    \@A \,\@X=\@B
\end{equation}
where
\begin{equation}
\@B=(\@I_2,\@I_2,\dots,\@I_2)^T,\quad \@X=\boldsymbol{\chi}^T,\quad \@A=\@I_{4J}-\left[(\@A_{j\ell}^T)_{1\leq j,\ell \leq 2J}\right].
    %\quad \textbf{I}=\text{diag}(\1,\1\dots,\1).
\end{equation}
Note that $\@A$ is a $4J\times 4J$ matrix (since each $\@A_{j\ell}$ is a $2\times 2$ matrix), and $\@X$ and $\@B$ are $4J\times 2$ matrices.
We now denote by $\@X^{(1)}$ and $\@X^{(2)}$ the first and second columns of $\@X$, respectively, and we use the same notation for $\@B$. Thus, the system \eqref{main_system} can now be split into two equations
\begin{equation}
\@A\,\@X^{(1)}=\@B^{(1)},\qquad \@A\,\@X^{(2)}=\@B^{(2)}.
\end{equation}
Using Cramer's rule, the solutions to these systems are
\begin{equation}
X^{(1,\ell)}=\frac{\det \@A^{(1,\ell)}_{r}}{\det \@A},\qquad X^{(2,\ell)}=\frac{\det \@A^{(2,\ell)}_{r}}{\det \@A},\quad \ell=1,\dots,4J,
\end{equation}
where
\begin{align*}
\@A^{(1,\ell)}_{r}=(\@A_1,\@A_2,\dots,\@A_{\ell-1},\@B^{(1)},\@A_{\ell+1},\dots,\@A_{4J}),\\
\@A^{(2,\ell)}_{r}=(\@A_1,\@A_2,\dots,\@A_{\ell-1},\@B^{(2)},\@A_{\ell+1},\dots,\@A_{4J}),
\end{align*}
and $\@A_1,\dots, \@A_{4J}$ are columns of $\@A$.
Note that $X^{(1,\ell)}$ and $X^{(2,\ell)}$ represent the $\ell$-th component of $\@X^{(1)}$ and $\@X^{(2)}$, respectively.
Finally, we have
\bse
\begin{equation}
\bar{\@N}^{\prime \, \textrm{(up)}}_n(\bar{z}_j)=\@X_j^T=\frac{1}{\det \@A}\begin{pmatrix}
        \det \@A^{(1,2j-1)}_{r} & \det \@A^{(1,2j)}_{r} \\
        \det \@A^{(2,2j-1)}_{r} & \det \@A^{(2,2j)}_{r}
    \end{pmatrix},\quad j=1,\dots,J,
\end{equation}
and
\begin{equation}
\bar{\@N}^{\prime  \, \textrm{(up)}}_n(i/z_j)=\@X_{j+J}^T=\frac{1}{\det \@A}\begin{pmatrix}
\det \@A^{(1,2(j+J)-1)}_{r} & \det \@A^{(1,2(j+J))}_{r}\\
\det \@A^{(2,2(j+J)-1)}_{r} & \det \@A^{(2,2(j+J))}_{r}
\end{pmatrix},\quad j=1,\dots, J.
\end{equation}
\ese
A generic multi-soliton solution can then be obtained by substituting the above expressions for the eigenfunctions in Eq.~\eqref{potential}.
\subsection{Fundamental soliton and breather interactions}
\begin{figure}[ht!]
\centering
    \includegraphics[width=.7\textwidth]{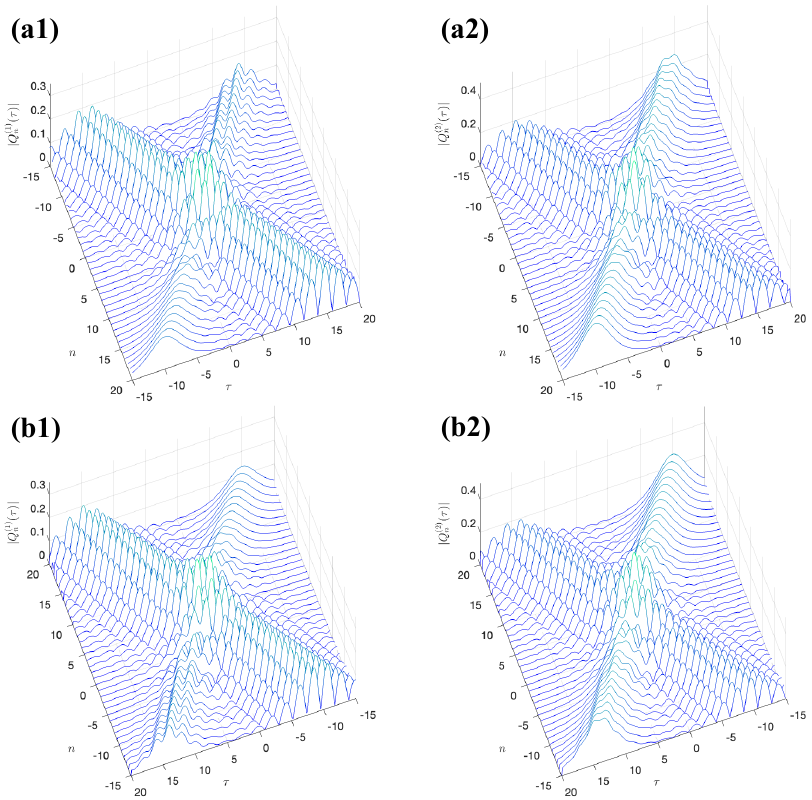}
    \caption{\textbf{(a)} Fundamental soliton-fundamental breather interaction with $z_{1}=\exp(0.15+i\pi/8)$, $\boldsymbol\gamma_{1}=(1,2)^{T}$, $\boldsymbol\delta_{1}=(1,0)^{T}$ and $z_{2}=\exp(0.1-i\pi/8)$, $\boldsymbol\gamma_{2}=(1,1)^{T}$, $\boldsymbol\delta_{2}=(0.1,0.1)^{T}$. \textbf{(b)} Reverse view, from which one can see more clearly that the fundamental soliton becomes a fundamental breather after the interaction.}
    \label{f:2soliton}
\end{figure}

\begin{figure}[ht!]
\centering
    \includegraphics[width=.8\textwidth]{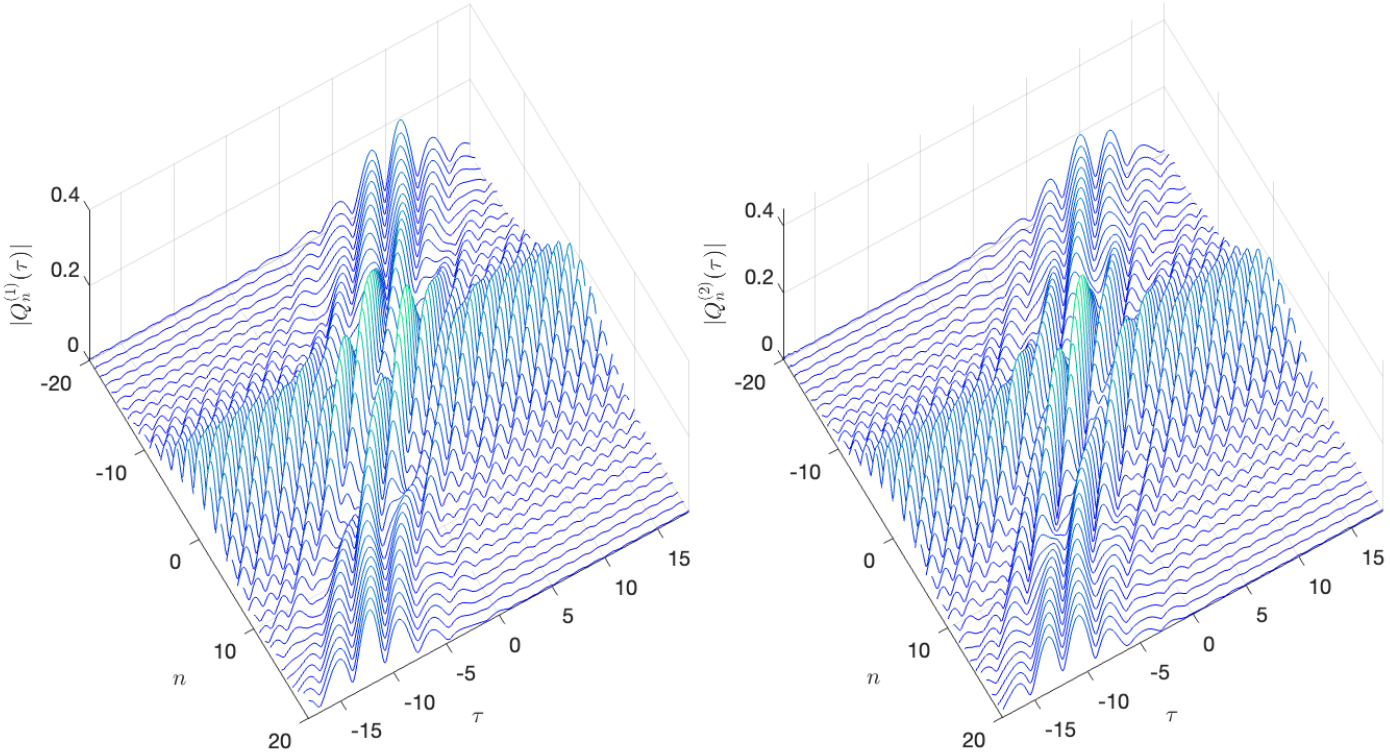}
    \caption{Interaction between two fundamental breathers with $z_{1}=\exp(0.1+i\pi/3)$, $\boldsymbol\gamma_{1}=(2,3)^{T}$, $\boldsymbol\delta_{1}=(0.04,0.04)^{T}$ and $z_{2}=\exp(0.12)$, $\boldsymbol\gamma_{2}=(0.2,0.2)^{T}$, $\boldsymbol\delta_{2}=(1,0.5)^{T}$.}
        \label{f:2breather}
\end{figure}

Presently, we focus on the case of $J=2$ fundamental breathers, whose norming constants are both of the form \eqref{e:Cfb}, in which case any terms with the products $\hat{\@C}_{j}\@C_{j}$ and $\bar{\@C}_{j}\tilde{\@C}_{j}$ vanish.  The case when one fundamental breather is a fundamental soliton can be obtained as a subcase (by assuming, as before, that one of $\boldsymbol{\delta}_j$ has only one non-zero component). In a similar way one can obtain the reduction to the formulas for 2-fundamental solitons, already considered in \cite{APT2004}.
Examples of soliton-breather and breather-breather solutions are plotted in Fig.~\ref{f:2soliton} and Fig.~\ref{f:2breather}, respectively. 
The explicit expressions of the solutions are too complicated to be of practical use ``as is'' even in these cases, but one can turn to long-time asymptotics to gain valuable insight into the behavior of the solutions. In particular, we aim to characterize the polarizations $\boldsymbol\gamma_{j}^{\pm}$, $\boldsymbol\delta_{j}^{\pm}$ (and in turn $\@u_{j}^{\pm}$, $\@v_{j}^{\pm}$) of both breathers as $\tau\rightarrow\pm\infty$. 

Writing $z_{j}=\exp(a_{j}+ib_{j})$, $\bar{z}_{j}=\exp(-a_{j}+ib_{j})$ with $a_j>0,\,b_j\in \Real$, and introducing
\begin{equation}
\label{xi}
\zeta_{j}=2a_{j}(n-v_{j}\tau)\,,\;\;v_{j}=-\frac{1}{a_{j}}\sinh2a_{j}\sin2b_{j}\,,\;\;\omega_{j}=\cosh2a_{j}\cos2b_{j}\,,\,\,j=1,2\,,
\end{equation}
together with the time evolution of the norming constants one can obtain the following useful expressions:
\bse
\label{Cs_xi}
\begin{eqnarray}
    \label{C_xi}
    z_{j}^{-2n}\@C_{j}(\tau)&=&\@C_{j}e^{-\zeta_{j}-2ib_{j}n+2i\omega_{j}\tau}\,,\\
    \label{Cbar_xi}
    \bar{z}_{j}^{2n}\bar{\@C}_{j}(\tau)&=&\bar{\@C}_{j}e^{-\zeta_{j}+2ib_{j}n-2i\omega_{j}\tau}\,,\\
    \label{Chat_xi}
     {z}_{j}^{-2n} \hat{\@C}_{j}(\tau)&=&\hat{\@C}_{j}e^{-\zeta_{j}-2ib_{j}n+2i\omega_{j}\tau}\,,\\
     \label{Ctilde_xi}
    \bar z_{j}^{2n}\tilde{\@C}_{j}(\tau)&=&\tilde{\@C}_{j}e^{-\zeta_{j}+2ib_{j}n-2i\omega_{j}\tau}\,,
\end{eqnarray}
\ese
where in the right-hand side $\@C_j$ denotes $\@C_j(0)$, and similarly for the other norming constants.
Without loss of generality, assume that $v_{1}<v_{2}$. In the reference frame of fundamental breather ``1'', i.e. for fixed $\zeta_{1}$, one can write
\begin{equation}
\label{e:zeta1vszeta2}
\zeta_{2}=\frac{a_{2}}{a_{1}}\zeta_{1}-2a_{2}(v_{2}-v_{1})\tau\,,
\end{equation}
from which it can be seen that $e^{-\zeta_{2}}\rightarrow0$ as $\tau\rightarrow-\infty$. Thus, in this limit, all terms in the linear system that contain a norming constant labeled ``2" can be neglected, in light of \eqref{Cs_xi} and \eqref{e:zeta1vszeta2}. The solution of the resulting asymptotic system in this case is straightforward, and it is given in detail in Appendix~B. The result is that as $\tau\rightarrow-\infty$ with $\zeta_{1}$ fixed, the 2-fundamental breather solution approaches a 1-fundamental breather identical to \eqref{fb}. Similarly, in the reference frame of fundamental breather ``2", one can express
\begin{equation}
    \zeta_{1}=\frac{a_{1}}{a_{2}}\zeta_{2}-2a_{1}(v_{1}-v_{2})\tau\,.
\end{equation}
From this, $e^{-\zeta_{1}}\rightarrow0$ as $\tau\rightarrow+\infty$, in which case all terms in the linear system that contain a norming constant labeled ``1" can be neglected. In this limit, one finds that the 2-fundamental breather solution again approaches a form identical to \eqref{fb}, but with parameters labeled ``2". A conclusion can then be drawn that the prescribed polarization vectors $\boldsymbol\gamma_{j}$ and $\boldsymbol\delta_{j}$ can be attributed to long time limits in the following way:
\begin{equation}
\label{pm}
\boldsymbol\gamma_{1}^{-}=\boldsymbol\gamma_{1}\,,\quad \boldsymbol\delta_{1}^{-}=\boldsymbol\delta_{1}\,, \quad
\boldsymbol\gamma_{2}^{+}=\boldsymbol\gamma_{2}\,,\quad \boldsymbol\delta_{2}^{+}=\boldsymbol\delta_{2}\,.
\end{equation}
\subsection{Explicit polarization maps}
Determining the long-time asymptotic behavior in the opposite limits (i.e., as $\tau \to +\infty$ with fixed $\zeta_1$, and as $\tau \to -\infty$ with fixed $\zeta_2$) and identifying the corresponding polarization vectors $\boldsymbol\gamma_{1}^{+}, \boldsymbol\delta_{1}^{+}$
and $\boldsymbol\gamma_{2}^{-}, \boldsymbol\delta_{2}^{-}$
is significantly more complicated, as one now has growing exponential terms, and at least 2 next-to-leading order terms have to be computed and retained at each step, because the system becomes degenerate. The detailed calculation of the asymptotics as $\tau \to -\infty$ with fixed $\zeta_2$ in the simpler case in which both interacting solitons are fundamental solitons is given in Appendix~B. In the more general case, in which one or both is a fundamental breather,
rather than deriving the polarization vectors directly from the long-time asymptotics, we assume:
\bse
\label{e:gamma+delta-}
\begin{eqnarray}
\label{gamma1+}
    \boldsymbol{\gamma}_{1}^{+}&=&\@c_{2}(z_{1},\boldsymbol{\gamma}_{2}^{+})\boldsymbol{\gamma}_{1}^{-}\,,\\
    \label{gamma2-}
    \boldsymbol{\gamma}_{2}^{-}&=&\@c_{1}(z_{2},\boldsymbol{\gamma}_{1}^{-})\boldsymbol{\gamma}_{2}^{+}\,,\\
    \label{delta1+}
    \boldsymbol{\delta}_{1}^{+}&=&
[\@a_2(z_{1},\boldsymbol{\delta}_{2}^{+})]^{\dagger}
\boldsymbol{\delta}_{1}^{-}
\equiv z_{2}^{2}\bar{z}_{2}^{-2}\bar{\@c}_{2}(1/z_{1}^{*},\boldsymbol{\delta}_{2}^{+})
    \boldsymbol{\delta}_{1}^{-}\,,\\
    \label{delta2-}
    \boldsymbol{\delta}_{2}^{-}&=&[\@a_{1}(z_{2},\boldsymbol{\delta}_{1}^{-})]^{\dagger}\boldsymbol{\delta}_{2}^{+}
\equiv z_{1}^{2}\bar{z}_{1}^{-2}\bar{\@c}_{1}(1/z_{2}^{*},\boldsymbol{\delta}_{1}^{-})\boldsymbol{\delta}_{2}^{+}\,,
\end{eqnarray}
\ese
where $\@c_j(z,\boldsymbol{\gamma}_j)$ and  $\bar{\@c}_j(z,\boldsymbol{\delta}_j)$ are the transmission coefficients associated to the $j$-th fundamental breather as given in \eqref{e:FB_transm}, and we have used
%Keeping in mind that
\begin{equation}
\@a_j(z,\boldsymbol{\delta}_{j})=z_{j}^{2}\bar{z}_{j}^{-2}\bar{\@c}_{j}^{\dagger}(1/z^{*},\boldsymbol{\delta}_{j}),
\end{equation}
on account of symmetry \eqref{ac_sym}, and $\Delta_\infty=\prod_{-\infty}^\infty (1+\alpha_n)=z_{j}^{2}\bar{z}_{j}^{-2}$ for any of the 1-soliton solutions with eigenvalue $z_j$.
Eqs.~\eqref{e:gamma+delta-} generalize Manakov's method to the fundamental breather case. 
%{\color{blue}(59c-d) were wrong due to a mistake in the code I used to verify the asymptotics. In this case, these results actually do work as a generalization of Manakov's method. I have changed the affected equations: (59c-d), (62b), (67c-d), (68), (71), (72c-d), (74).}
Indeed, if we let $\@C_1^\pm=\boldsymbol{\gamma}_1^\pm \left(\boldsymbol{\delta}_1^\pm\right)^\dagger$, then Eqs.~\eqref{e:gamma+delta-} give
$$
\@C_1^+=\@c_2(z_1,\boldsymbol{\gamma_2}^+)\@C_1^- \@a_2(z_1,\boldsymbol{\delta}_2^+),
\qquad
\@C_2^-=\@c_1(z_2,\boldsymbol{\gamma_1}^-)\@C_2^+ \@a_1(z_2,\boldsymbol{\delta}_1^-),
$$
which, upon identifying $\@S_j^\pm \leftrightarrow \@C_j^\pm$ for $j=1,2$, reduce to Eqs.~(5.3.208) in \cite{APT2004} in the case of fundamental solitons, i.e., when $\@a_j(z)$ is independent of $\boldsymbol{\delta}_j^\pm$ and diagonal. Moreover, Eqs.~\eqref{e:gamma+delta-} are also consistent with the direct computation of the long-time asymptotics in Appendix~B.
Similar equations were obtained in \cite{CGP23} for the ccSPE using the dressing method
and the Darboux matrices for fundamental breather solutions. Here, we will, on one hand, verify numerically that \eqref{e:gamma+delta-} provide the correct long-time asymptotics for the soliton-breather and breather-breather interaction, and, on the other hand, show how the maps defining the polarization shifts that follow from \eqref{e:gamma+delta-} can be derived from the refactorization property of the associated transmission coefficients. 
%The numerical checks for various interaction scenarios are shown in Figs.~\ref{f:2soliton-asymp}, \ref{f:2breather-asymp}, \ref{f:fs-cb-asymp}.

Specifically, we can write the transmission coefficients \eqref{e:FB_transm} as follows:
\bse
\label{FB_coef}
\begin{eqnarray}
\@c_{j}(z,\boldsymbol{\gamma}_{j})=\alpha_{j}(z)\left[\@I_{2}+\beta_{j}(z)\frac{\boldsymbol{\gamma}_{j}\boldsymbol{\gamma}_{j}^{\dagger}}{\boldsymbol{\gamma}_{j}^{\dagger}\boldsymbol{\gamma}_{j}}\right], \qquad
\bar{\@c}_{j}(z,\boldsymbol{\delta}_{j})=\bar\alpha_{j}(z)\left[\@I_{2}+\bar\beta_{j}(z)\frac{\boldsymbol{\delta}_{j}\boldsymbol{\delta}_{j}^{\dagger}}{\boldsymbol{\delta}_{j}^{\dagger}\boldsymbol{\delta}_{j}}\right],
\end{eqnarray}
with
\begin{eqnarray}
    &\displaystyle\alpha_{j}(z)=\frac{\bar{z}_{j}^{2}+z^{-2}}{z_{j}^{2}+z^{-2}},\;\;\;&\displaystyle\beta_{j}(z)=\frac{(\bar{z}_{j}^{2}-z_{j}^{2})(\bar{z}_{j}^{2}+z_{j}^{-2})}{(z^{2}-\bar{z}_{j}^{2})(\bar{z}_{j}^{2}+z^{-2})},\\
    &\bar\alpha_{j}(z)=\alpha_{j}^{*}(1/z^{*}),\;\;\;&\bar\beta_{j}(z)=\beta_{j}^{*}(1/z^{*}).
\end{eqnarray}
\ese
Alternatively, one can express the polarization shifts in terms of the unit vectors $\@u_{j}^{\pm}=(\boldsymbol\gamma_{j}^{\pm})^*/\|\boldsymbol\gamma_{j}^{\pm}\|$ and $\@v_{j}^{\pm}=(\boldsymbol\delta_{j}^{\pm})^*/\|\boldsymbol\delta_{j}^{\pm}\|$ as follows: 
\bse
\label{e:u+v-}
\begin{eqnarray}
&\displaystyle\@u_{1}^{+}=\frac{\|\boldsymbol\gamma_{1}^{-}\|}{\|\boldsymbol\gamma_{1}^{+}\|}[\@c_{2}(z_{1},\@u_{2}^{+})]^{*}\@u_{1}^{-}\,,\;\;\;
    &\@u_{2}^{-}=\frac{\|\boldsymbol\gamma_{2}^{+}\|}{\|\boldsymbol\gamma_{2}^{-}\|}[\@c_{1}(z_{2},\@u_{1}^{-})]^{*}\@u_{2}^{+}\,,\\
    &\displaystyle\@v_{1}^{+}=
\frac{\|\boldsymbol\delta_{1}^{-}\|}{\|\boldsymbol\delta_{1}^{+}\|}[\@a_2(z_{1},\@v_{2}^{+})]^{T}
\@v_{1}^{-}\,,\;\;\;
   &\@v_{2}^{-}=\frac{\|\boldsymbol\delta_{2}^{+}\|}{\|\boldsymbol\delta_{2}^{-}\|}[\@a_{1}(z_{2},\@v_{1}^{-})]^{T}\@v_{2}^{+}\,.
\end{eqnarray}
\ese
One can check that%, in agreement with \cite{APT06},
\begin{equation}
\frac{\|\boldsymbol\gamma_{1}^{-}\|\|\boldsymbol\delta_{1}^{-}\|}{\|\boldsymbol\gamma_{1}^{+}\|\|\boldsymbol\delta_{1}^{+}\|}=\frac{\|\boldsymbol\gamma_{2}^{+}\|\|\boldsymbol\delta_{2}^{+}\|}{\|\boldsymbol\gamma_{2}^{-}\|\|\boldsymbol\delta_{2}^{-}\|}=:\chi\,,
\end{equation}
and direct calculations also show that
\begin{equation}
    \frac{z_{1}}{\bar{z}_{1}}\frac{\|\boldsymbol\gamma_{1}^{-}\|}{\|\boldsymbol\gamma_{1}^{+}\|}=\frac{z_{2}}{\bar{z}_{2}}\frac{\|\boldsymbol\gamma_{2}^{+}\|}{\|\boldsymbol\gamma_{2}^{-}\|}=:\chi_{\gamma}\,,\;\;\;\;\;\;\frac{\bar{z}_{1}}{z_{1}}\frac{\|\boldsymbol\delta_{1}^{-}\|}{\|\boldsymbol\delta_{1}^{+}\|}=\frac{\bar{z}_{2}}{z_{2}}\frac{\|\boldsymbol\delta_{2}^{+}\|}{\|\boldsymbol\delta_{2}^{-}\|}=:\chi_{\delta}\,,
\end{equation}
with 
\bse
\label{chis}
\begin{eqnarray}
    \chi_{\gamma}^{2}&=&\frac{z_{2}^{2}}{\bar{z}_{2}^{2}}\bigg\vert\frac{\bar{z}_{1}^{-2}+\bar{z}_{2}^{2}}{z_{1}^{-2}+\bar{z}_{2}^{2}}\bigg\vert^{2}\bigg[1+\frac{(z_{1}^{2}-\bar{z}_{1}^{2})(z_{1}^{-2}+\bar{z}_{1}^{2})(z_{2}^{2}-\bar{z}_{2}^{2})(z_{2}^{2}+\bar{z}_{2}^{-2})}{(\bar{z}_{1}^{2}-\bar{z}_{2}^{2})(\bar{z}_{2}^{-2}+\bar{z}_{1}^{2})(z_{2}^{2}-z_{1}^{2})(z_{2}^{2}+z_{1}^{-2})}\big\vert\@u_{1}^{-\,\dagger}\@u_{2}^{-}\big\vert^{2}\bigg]\,,\\
    \chi_{\delta}^{2}&=&\frac{\bar z_{2}^{2}}{{z}_{2}^{2}}\bigg\vert\frac{{z}_{1}^{-2}+{z}_{2}^{2}}{\bar z_{1}^{-2}+{z}_{2}^{2}}\bigg\vert^{2}\bigg[1+\frac{(z_{1}^{2}-\bar{z}_{1}^{2})(z_{1}^{-2}+\bar{z}_{1}^{2})(z_{2}^{2}-\bar{z}_{2}^{2})(z_{2}^{2}+\bar{z}_{2}^{-2})}{(\bar{z}_{1}^{2}-\bar{z}_{2}^{2})(\bar{z}_{2}^{-2}+\bar{z}_{1}^{2})(z_{2}^{2}-z_{1}^{2})(z_{2}^{2}+z_{1}^{-2})}\big\vert\@v_{1}^{-\,\dagger}\@v_{2}^{-}\big\vert^{2}\bigg]\,.
\end{eqnarray}
\ese
With the help of these definitions, the polarization shifts \eqref{e:u+v-} can be written explicitly as
\bse
\label{e:u+v-explicit}
\begin{eqnarray}
\label{u1+}
    \displaystyle\@u_{1}^{+}&=&\chi_{\gamma}\frac{\bar{z}_{1}}{z_{1}}\frac{{z}_{2}^{-2}+\bar z_{1}^{2}}{\bar z_{2}^{-2}+\bar z_{1}^{2}}\bigg[\@u_{1}^{-}+\frac{({z}_{2}^{-2}-\bar z_{2}^{-2})({z}_{2}^{-2}+\bar z_{2}^{2})}{(\bar z_{1}^{-2}-{z}_{2}^{-2})({z}_{2}^{-2}+\bar z_{1}^{2})}\big(\@u_{2}^{+\,\dagger}\@u_{1}^{-}\big)\@u_{2}^{+}\bigg]\,,\\
\label{u2-}
    \displaystyle\@u_{2}^{-}&=&\chi_{\gamma}\frac{\bar{z}_{2}}{z_{2}}\frac{{z}_{1}^{-2}+\bar z_{2}^{2}}{\bar z_{1}^{-2}+\bar z_{2}^{2}}\bigg[\@u_{2}^{+}+\frac{({z}_{1}^{-2}-\bar z_{1}^{-2})({z}_{1}^{-2}+\bar z_{1}^{2})}{(\bar z_{2}^{-2}-{z}_{1}^{-2})({z}_{1}^{-2}+\bar z_{2}^{2})}\big(\@u_{1}^{-\,\dagger}\@u_{2}^{+}\big)\@u_{1}^{-}\bigg]\,,\\
\label{v1+}
    \displaystyle\@v_{1}^{+}&=&\chi_{\delta}\frac{{z}_{1}}{\bar z_{1}}\frac{\bar{z}_{2}^{-2}+ z_{1}^{2}}{z_{2}^{-2}+z_{1}^{2}}\bigg[\@v_{1}^{-}+\frac{(\bar{z}_{2}^{2}-z_{2}^{2})(\bar{z}_{2}^{2}+z_{2}^{-2})}{( z_{1}^{2}-\bar{z}_{2}^{2})(\bar{z}_{2}^{2}+z_{1}^{-2})}\big(\@v_{2}^{+\,\dagger}\@v_{1}^{-}\big)\@v_{2}^{+}\bigg]\,,\\
\label{v2-}
    \displaystyle\@v_{2}^{-}&=&\chi_{\delta}\frac{{z}_{2}}{\bar z_{2}}\frac{\bar{z}_{1}^{-2}+z_{2}^{2}}{ z_{1}^{-2}+ z_{2}^{2}}\bigg[\@v_{2}^{+}+\frac{(\bar{z}_{1}^{2}-z_{1}^{2})(\bar{z}_{1}^{2}+z_{1}^{-2})}{( z_{2}^{2}-\bar{z}_{1}^{2})(\bar{z}_{1}^{2}+ z_{2}^{-2})}\big(\@v_{1}^{-\,\dagger}\@v_{2}^{+}\big)\@v_{1}^{-}\bigg]\,.
\end{eqnarray}
\ese
One can verify that the above formulas reduce to known results for the interaction of 2 fundamental solitons. For instance, by taking $\@v_{1}^{-}=\@v_{2}^{+}=(1,0)^{T}$ and $\@v_{1}^{+}=(\nu^{+},0)^{T}$, \eqref{u1+} and \eqref{v1+} can be reduced to Eq.~(5.3.217) of \cite{APT2004}. Particularly, from \eqref{v1+} it can be found that
\begin{equation}
    \nu^{+}=
\chi_{\delta}\frac{z_1}{\bar{z}_{1}}\frac{z_{1}^{2}-{z}_{2}^{2}}{{z}_{1}^{2}-\bar z_{2}^{2}}\,.
\end{equation}
Then, by rewriting \eqref{u1+} for $\@p_{1}^{+}\equiv\@u_{1}^{+}(\nu^{+})^*$ (with $\@p_{1}^{-}\equiv\@u_{1}^{-}$ and $\@p_{2}^{+}\equiv\@u_{2}^{+}$) and using $\chi_{\gamma}\chi_{\delta}=\chi$, one arrives at (5.3.217) in \cite{APT2004}, which in turn matches the result from the long-time asymptotics in Appendix~B. 

In their current form \eqref{e:u+v-explicit}, the polarization shifts are expressed as maps $(\@u_{1}^{-},\@u_{2}^{+})
\mapsto(\@u_{1}^{+},\@u_{2}^{-})$ 
and $(\@v_{1}^{-},\@v_{2}^{+})\mapsto (\@v_{1}^{+},\@v_{2}^{-})$
due to our knowledge of how the prescribed norming constants are attributed to the $\tau\rightarrow\pm\infty$ limits, see \eqref{pm}. We remark that one can instead view these as maps from the polarization vectors before the interaction to those after the interaction, i.e. $(\@u_{1}^{-},\@u_{2}^{-})\mapsto (\@u_{1}^{+},\@u_{2}^{+})$ and $(\@v_{1}^{-}, \@v_{2}^{-})\mapsto(\@v_{1}^{+},\@v_{2}^{+})$. In particular, observe that 
\begin{eqnarray}
    &&\@c_{j}(z,\boldsymbol\gamma_{j})^{-1}=z_{j}^{2}\bar{z}_{j}^{-2}\@c_{j}^{\dagger}(1/z^{*},\boldsymbol\gamma_{j})\,,\\
    &&\@a_{j}(z,\boldsymbol\delta_{j})^{-1}=z_{j}^{-2}\bar{z}_{j}^{2}\@a_{j}^{\dagger}(1/z^{*},\boldsymbol\delta_{j})\,,
\end{eqnarray}
which implies that we can rewrite \eqref{gamma2-} and \eqref{delta2-} as
\begin{equation}
    \boldsymbol\gamma_{2}^{+}=z_{1}^{2}\bar{z}_{1}^{-2}{\@c}_{1}^{\dagger}(1/z_{2}^{*},\boldsymbol\gamma_{1}^{-})\boldsymbol\gamma_{2}^{-}\,,\;\;\;\;\boldsymbol\delta_{2}^{+}=z_{1}^{-2}\bar{z}_{1}^{2}{\@a}_{1}(1/z_{2}^{*},\boldsymbol\delta_{1}^{-})\boldsymbol\delta_{2}^{-}\,.
\end{equation}
Then, the above can be substituted into \eqref{gamma1+} and \eqref{delta1+} so that all ``$+$" polarization vectors are expressed in terms of ``$-$" ones. After simplification, we have:
\bse
\label{e:u+v-new}
\begin{eqnarray}
  \label{e:u+v-new1}  \@u_{1}^{+}&=&\frac{1}{\chi_{\gamma}}\frac{{z}_{1}}{\bar z_{1}}\frac{{z}_{2}^{-2}+z_{1}^{2}}{\bar z_{2}^{-2}+z_{1}^{2}}\bigg[\@u_{1}^{-}+\frac{({z}_{2}^{-2}-\bar z_{2}^{-2})({z}_{2}^{-2}+\bar z_{2}^{2})}{( z_{1}^{-2}-{z}_{2}^{-2})({z}_{2}^{-2}+z_{1}^{2})}\big(\@u_{2}^{-\,\dagger}\@u_{1}^{-}\big)\@u_{2}^{-}\bigg]\,,\\
   \label{e:u+v-new2} \@u_{2}^{+}&=&\frac{1}{\chi_{\gamma}}\frac{{z}_{2}}{\bar z_{2}}\frac{z_{2}^{2}+\bar z_{1}^{-2}}{z_{2}^{2}+{z}_{1}^{-2}}\bigg[\@u_{2}^{-}+\frac{({z}_{1}^{-2}+\bar z_{1}^{2})({z}_{1}^{2}-\bar z_{1}^{2})}{(\bar z_{1}^{2}-z_{2}^{2})(z_{2}^{-2}+\bar z_{1}^{2})}\big(\@u_{1}^{-\,\dagger}\@u_{2}^{-}\big)\@u_{1}^{-}\bigg]\,,\\
    \@v_{1}^{+}&=&\frac{1}{\chi_{\delta}}\frac{\bar{z}_{1}}{ z_{1}}\frac{\bar{z}_{2}^{-2}+ \bar z_{1}^{2}}{z_{2}^{-2}+ \bar z_{1}^{2}}\bigg[\@v_{1}^{-}+\frac{(\bar{z}_{2}^{2}- z_{2}^{2})(\bar{z}_{2}^{2}+ z_{2}^{-2})}{(\bar z_{1}^{2}-\bar{z}_{2}^{2})(\bar{z}_{2}^{2}+\bar z_{1}^{-2})}\big(\@v_{2}^{-\,\dagger}\@v_{1}^{-}\big)\@v_{2}^{-}\bigg]\,,\\
    \@v_{2}^{+}&=&\frac{1}{\chi_{\delta}}\frac{\bar{z}_{2}}{z_{2}}\frac{\bar z_{2}^{2}+z_{1}^{-2}}{ \bar z_{2}^{2}+\bar{z}_{1}^{-2}}\bigg[\@v_{2}^{-}+\frac{(\bar{z}_{1}^{2}+ z_{1}^{-2})(\bar{z}_{1}^{-2}-z_{1}^{-2})}{(z_{1}^{-2}-\bar z_{2}^{-2})(\bar z_{2}^{2}+z_{1}^{-2})}\big(\@v_{1}^{-\,\dagger}\@v_{2}^{-}\big)\@v_{1}^{-}\bigg]\,.
\end{eqnarray}
\ese
Note that from the form given in \eqref{e:u+v-explicit}, we see that the shifts are symmetric with respect to interchanging $1\leftrightarrow2$ and $+\leftrightarrow-$. On the other hand, \eqref{e:u+v-new} are no longer symmetric upon interchanging $1\leftrightarrow2$, consistent with what one observes in the Manakov formulas \eqref{e:Manakov_p}. 
%Mathematica has verified (21) and (22) symbolically. 
%{\color{red}Is there a way we can deduce the polarization shift without the prior knowledge of the ccSP system and without Darboux matrices?}

\begin{figure}[ht!]
\centering
    \includegraphics[width=\textwidth]{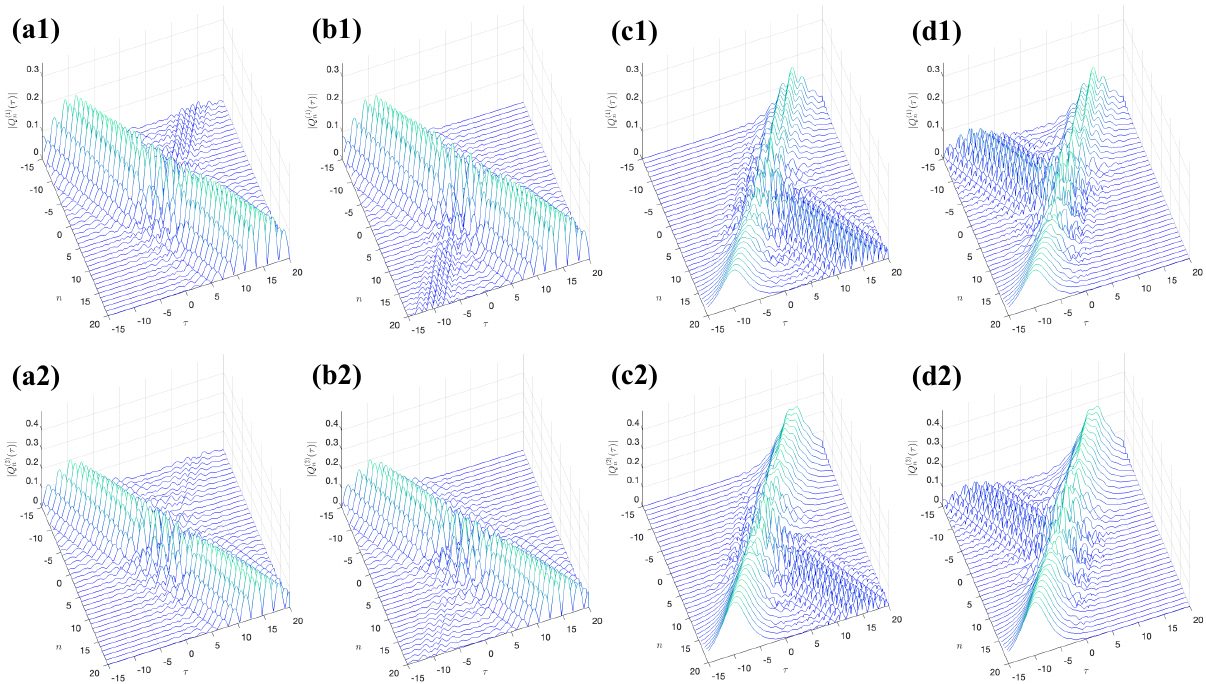}
    \caption{The same soliton-breather interaction as in Fig.~\ref{f:2soliton}, with the predicted asymptotic breathers subtracted in each direction. In particular; in \textbf{(a)} the soliton with polarization vectors $\@u_{1}^{-},\@v_{1}^{-}$ is subtracted, in \textbf{(b)} the breather with polarization vectors $\@u_{1}^{+},\@v_{1}^{+}$ is subtracted, in \textbf{(c)} the breather with polarization vectors $\@u_{2}^{-},\@v_{2}^{-}$ is subtracted, in \textbf{(d)} the breather with polarization vectors $\@u_{2}^{+},\@v_{2}^{+}$ is subtracted.}
    \label{f:2soliton-asymp}
\end{figure}

\begin{figure}[ht!]
\centering
    \includegraphics[width=\textwidth]{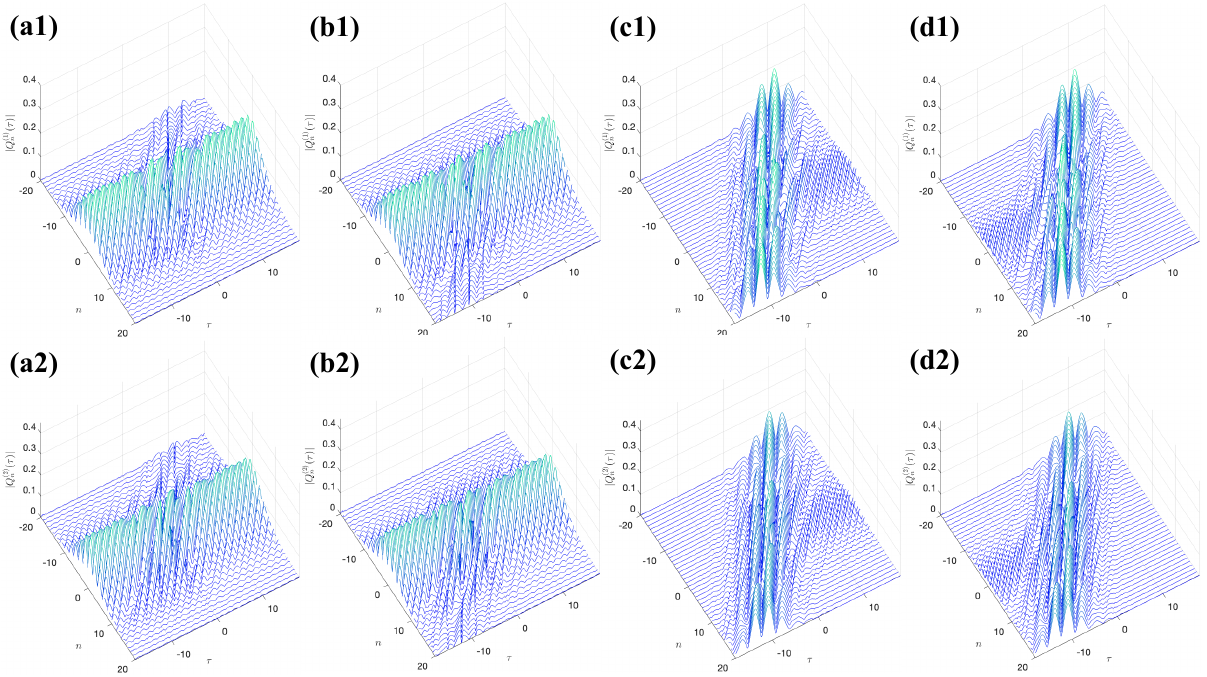}
    \caption{The same breather-breather interaction as in Fig.~\ref{f:2breather}, with the predicted asymptotic breathers subtracted in each direction. In particular; in \textbf{(a)} the breather with polarization vectors $\@u_{1}^{-},\@v_{1}^{-}$ is subtracted, in \textbf{(b)} the breather with polarization vectors $\@u_{1}^{+},\@v_{1}^{+}$ is subtracted, in \textbf{(c)} the breather with polarization vectors $\@u_{2}^{-},\@v_{2}^{-}$ is subtracted, in \textbf{(d)} the breather with polarization vectors $\@u_{2}^{+},\@v_{2}^{+}$ is subtracted.}
    \label{f:2breather-asymp}
\end{figure}

Fig.~\ref{f:2soliton-asymp} shows the same soliton-breather solution as in Fig.~\ref{f:2soliton}, but with the long-time asymptotics predicted by \eqref{pm} and \eqref{e:u+v-explicit} subtracted out. Similarly, Fig.~\ref{f:2breather-asymp}  shows the 2-fundamental breather solution as in Fig.~\ref{f:2breather} with the asymptotics subtracted out, providing numerical verifications of the correctness of the long-time asymptotics obtained from \eqref{e:gamma+delta-}.


\paragraph{Fundamental soliton-fundamental breather interaction.}
Consider the case where soliton 1 is a fundamental soliton before the interaction, i.e.
\begin{equation}
\@v_{1}^{-}=(1,0)^{T}, \qquad \@v_{2}^{+}=(\mu,\kappa)^{T}.
\end{equation}
Then, according to \eqref{e:gamma+delta-}, after the interaction we have:
\begin{equation}
\displaystyle\@v_{1}^{+}=\chi_{\delta}\frac{{z}_{1}}{\bar z_{1}}\frac{\bar{z}_{2}^{-2}+ z_{1}^{2}}{z_{2}^{-2}+z_{1}^{2}}\bigg[\begin{pmatrix}1\\0\end{pmatrix}+\frac{(\bar{z}_{2}^{2}-z_{2}^{2})(\bar{z}_{2}^{2}+z_{2}^{-2})}{( z_{1}^{2}-\bar{z}_{2}^{2})(\bar{z}_{2}^{2}+z_{1}^{-2})}\mu^{*}\begin{pmatrix}\mu\\\kappa\end{pmatrix}\bigg]\,.
\end{equation}
If $\mu$ and $\kappa$ are both nonzero, both components of $\@v_1^+$ from the above equation are generically nonzero, so soliton 1 becomes itself a fundamental breather upon interaction with the fundamental breather. This phenomenon can be observed in Fig.~\ref{f:2soliton}.

%{\color{red}
%If we solve the issue with the adjoint/complex conjugate in the Manakov's interpretation of the method, we can probably describe other types of interactions, see below.
\subsection{Interactions involving composite breathers}
\begin{figure}[ht!]
\centering
    \includegraphics[width=0.8\textwidth]{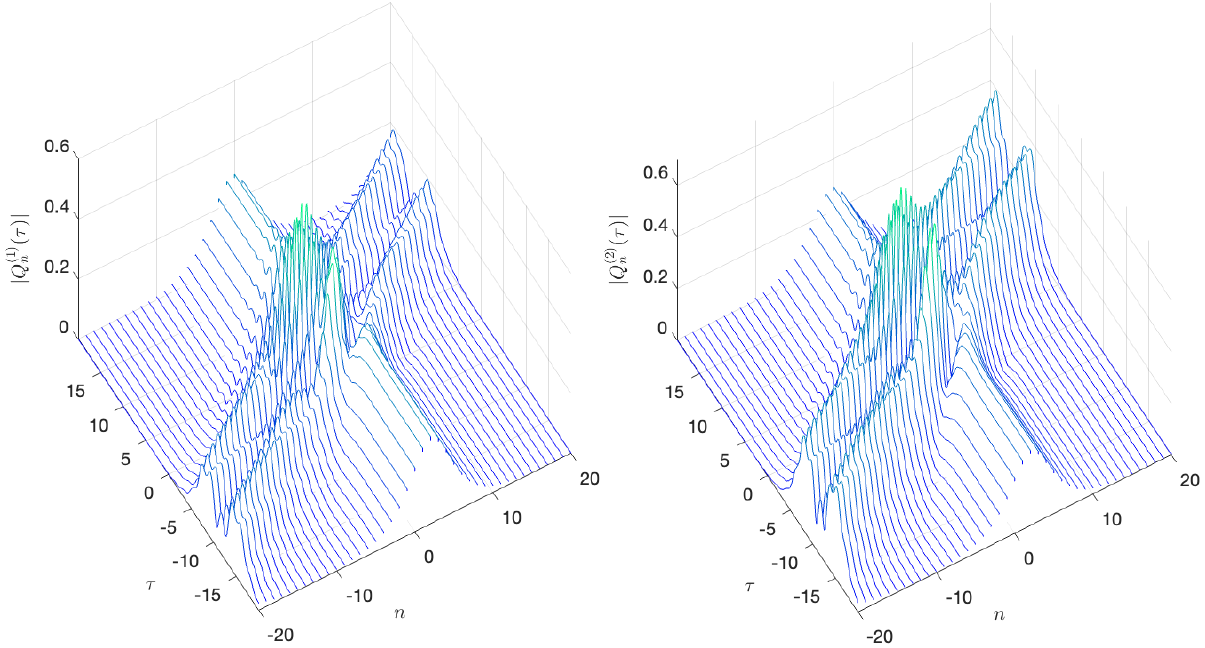}
    \caption{Interaction of a fundamental soliton with $z_{1}=\exp(0.2)$, $\boldsymbol\gamma_{1}=(1,1)^{T}$, $\boldsymbol\delta_{1}=(1,0)^{T}$ and a composite breather with $z_{2}=\exp(0.2-i\pi/8)$, $\boldsymbol\gamma_{2}=(1,2)^{T}$, $\boldsymbol\varepsilon_{2}=(0.1,0.1)^{T}$, $\boldsymbol\delta_{2}=(1,0)^{T}$, $\tilde{\boldsymbol\delta}_{2}=(0,1)^{T}$.}
    \label{f:fs-cb}
\end{figure}

\begin{figure}[ht!]
\centering
    \includegraphics[width=\textwidth]{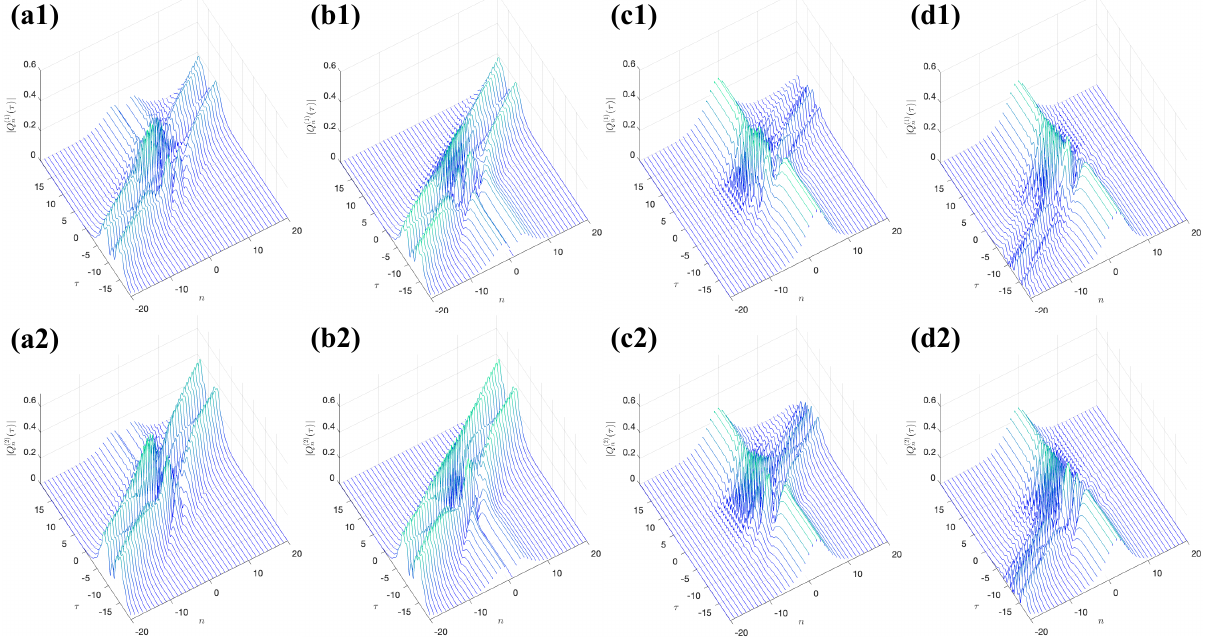}
    \caption{The same interaction as in Fig.~\ref{f:fs-cb}, with the predicted asymptotic states subtracted in each direction. In particular, in \textbf{(a)} the fundamental soliton with polarization vectors $\boldsymbol\gamma_{1}^{-},\boldsymbol\delta_{1}^{-}$ is subtracted, in \textbf{(b)} the fundamental soliton with polarization vectors $\boldsymbol\gamma_{1}^{+},\boldsymbol\delta_{1}^{+}$ is subtracted, in \textbf{(c)} the composite breather with polarization vectors $\boldsymbol\gamma_{2}^{-},\boldsymbol\varepsilon_{2}^{-},\boldsymbol\delta_{2}^{-},\tilde{\boldsymbol\delta}_{2}^{-}$ is subtracted, in \textbf{(d)} the composite breather with polarization vectors $\boldsymbol\gamma_{2}^{+},\boldsymbol\varepsilon_{2}^{+},\boldsymbol\delta_{2}^{+},\tilde{\boldsymbol\delta}_{2}^{+}$ is subtracted.}
    \label{f:fs-cb-asymp}
\end{figure}

\paragraph{Interactions of 2 composite breathers.}
Recalling that the norming constant of a composite breather is a full-rank $2\times 2$ matrix, we can separate the two columns by letting 
\begin{equation}
\label{e:cb_C}
\@C_1=\boldsymbol{\gamma}_1\boldsymbol{\delta}_1^\dagger +
\boldsymbol{\varepsilon}_1\tilde{\boldsymbol{\delta}}_1^\dagger,
\end{equation}
and choose $\boldsymbol{\delta}_1=(1,0)^T$,  $\tilde{\boldsymbol{\delta}}_1=(0,1)^T$, and $\boldsymbol{\gamma}_1$ and $\boldsymbol{\varepsilon}_1$ any two linearly independent vectors. Similarly to what was shown in the ccSPE \cite{CGP23}, we assume that $\boldsymbol{\varepsilon}_1$ transforms like $\boldsymbol{\gamma}_1$, and $\tilde{\boldsymbol{\delta}}_1$ transforms like $\boldsymbol{\delta}_1$ in \eqref{e:gamma+delta-}, with the appropriate transmission coefficients. Since the latter are all proportional to the identity in the composite breather case (cf Eqs.~\eqref{e:CB_transm}), this suffices to show that the interaction of 2 composite breathers is always trivial.
\paragraph{Interaction between a fundamental and a composite breather.}
Let us assume soliton 1 is either a fundamental soliton or a fundamental breather (i.e., with a rank-1 norming constant), while soliton 2 is a composite breather, i.e., we take
\begin{equation}
\@C_1=\boldsymbol{\gamma}_1\boldsymbol{\delta}_1^\dagger, \qquad
\@C_2=\boldsymbol{\gamma}_2\boldsymbol{\delta}_2^\dagger +
\boldsymbol{\varepsilon}_2\tilde{\boldsymbol{\delta}}_2^\dagger,
\end{equation}
where $\boldsymbol{\delta}_2=(1,0)^T$,  $\tilde{\boldsymbol{\delta}}_2=(0,1)^T$, and $\boldsymbol{\gamma}_2$ and $\boldsymbol{\varepsilon}_2$ are linearly independent vectors.
%When one composite breather interacts with either a fundamental soliton or a fundamental breather, the interaction would be\footnote{Cf (186) in the ccSPE paper}:
From Eqs.~\eqref{e:gamma+delta-} we have
\begin{gather}
\boldsymbol{\gamma}_1^+=\@c_2(z_1,\boldsymbol{\gamma}_2^+)
\boldsymbol{\gamma}_1^-, \qquad
\boldsymbol{\delta}_1^+=
[\@a_2(z_1,\boldsymbol{\delta}_2^+)]^\dagger
\boldsymbol{\delta}_1^-,
\end{gather}
%If we assume that soliton 1 is either a fundamental soliton or a fundamental breather, while soliton 2 is a composite breather, 
and since for the composite breather
$$
\@c_2(z_1,\boldsymbol{\gamma}_2^+)=\frac{(z_1^{-2}-z_2^{-2})(z_1^{-2}+\bar{z}_2^2)}{(z_1^{-2}-\bar{z}_2^{-2})(z_1^{-2}+z_2^2)}\@I_2, \qquad
\@a_2(z_1,\boldsymbol{\delta}_2^+)=\frac{(z_1^{2}-z_2^{2})(z_1^{-2}+\bar{z}_2^2)}{(z_1^{2}-\bar{z}_2^2)(z_1^{-2}+z_2^2)}\@I_2
$$
(cf. \eqref{e:CB_transm})
we have
\bse
\label{fs-cb-shifts1}
\begin{gather}
\boldsymbol{\gamma}_1^-=\boldsymbol{\gamma}_1, \qquad \boldsymbol{\delta}_1^-=\boldsymbol{\delta}_1, \\
\boldsymbol{\gamma}_1^+=\frac{(z_1^{-2}-z_2^{-2})(z_1^{-2}+\bar{z}_2^2)}{(z_1^{-2}-\bar{z}_2^{-2})(z_1^{-2}+z_2^2)}\boldsymbol{\gamma}_1^-, \qquad
\boldsymbol{\delta}_1^+=\frac{(\bar z_1^{-2}-\bar z_2^{-2})(\bar z_1^{2}+{z}_2^{-2})}{(\bar z_1^{-2}-{z}_2^{-2})(\bar z_1^{2}+\bar z_2^{-2})}\boldsymbol{\delta}_1^-,
\end{gather}
\ese
showing that the rank-1 soliton (be it a fundamental soliton or a fundamental breather) is essentially unaffected by the interaction with the composite breather, as the interaction only results in a shift in the overall phase and in the soliton center. For the composite breather, assuming again that like in the ccSPE $\boldsymbol{\varepsilon}_2$ transforms like $\boldsymbol{\gamma}_2$, and $\tilde{\boldsymbol{\delta}}_2$ transforms like $\boldsymbol{\delta}_2$ in \eqref{e:gamma+delta-}, with the appropriate transmission coefficients, we find
\bse
\label{fs-cb-shifts2}
\begin{gather}
\boldsymbol{\gamma}_2^+=\boldsymbol{\gamma}_2, \qquad
\boldsymbol{\delta}_2^+=\boldsymbol{\delta}_2\equiv \begin{pmatrix}
1 \\ 0    
\end{pmatrix}, \qquad
\boldsymbol{\varepsilon}_2^+=\boldsymbol{\varepsilon}_2, \qquad
\tilde{\boldsymbol{\delta}}_2^+=\tilde{\boldsymbol{\delta}}_2\equiv \begin{pmatrix}
0 \\ 1    
\end{pmatrix}, \\
\boldsymbol{\gamma}_2^-=\@c_1(z_2,\boldsymbol{\gamma}_1^-)
\boldsymbol{\gamma}_2^+\equiv 
\frac{\bar{z}_{1}^{2}+z_{2}^{-2}}{z_{1}^{2}+z_{2}^{-2}}\left[\@I_2 
+\frac{(\bar{z}_{1}^{2}-z_{1}^{2})(\bar{z}_{1}^{2}+z_{1}^{-2})}{(z_{2}^{2}-\bar{z}_{1}^{2})(\bar{z}_{1}^{2}+z_{2}^{-2})}
\frac{\boldsymbol{\gamma}_1^-(\boldsymbol{\gamma}_1^{-})^\dagger}{\|\boldsymbol{\gamma}_1^-\|^2}
\right]\boldsymbol{\gamma}_2^+,\\
\boldsymbol{\varepsilon}_2^-=\@c_1(z_2,\boldsymbol{\gamma}_1^-)\boldsymbol{\varepsilon}_2^+
\equiv \frac{\bar{z}_{1}^{2}+z_{2}^{-2}}{z_{1}^{2}+z_{2}^{-2}}\left[\@I_2 
+\frac{(\bar{z}_{1}^{2}-z_{1}^{2})(\bar{z}_{1}^{2}+z_{1}^{-2})}{(z_{2}^{2}-\bar{z}_{1}^{2})(\bar{z}_{1}^{2}+z_{2}^{-2})}
\frac{\boldsymbol{\gamma}_1^-(\boldsymbol{\gamma}_1^{-})^\dagger}{\|\boldsymbol{\gamma}_1^-\|^2}
\right]\boldsymbol{\varepsilon}_2^+,\\
\boldsymbol{\delta}_2^-=[\@a_1(z_2,\boldsymbol{\delta}_1^-)]^\dagger
\boldsymbol{\delta}_2^+\equiv 
\frac{ \bar z_2^{-2}+{z}_1^{2}}{ \bar z_2^{-2}+\bar z_1^{2}}\left[\@I_2 
+\frac{({z}_1^{-2}+ \bar z_1^{2})( \bar z_1^{2}-{z}_1^{-2})
}{({z}_1^{-2}- \bar z_2^{-2})( \bar z_2^{2}+{z}_1^{-2})}
\frac{\boldsymbol{\delta}_1^-(\boldsymbol{\delta}_1^{-})^\dagger}{\|\boldsymbol{\delta}_1^-\|^2}
\right]\boldsymbol{\delta}_2^+,
\\
\tilde{\boldsymbol{\delta}}_2^-=
[\@a_1(z_2,{\boldsymbol{\delta}}_1^-)]^\dagger\tilde{\boldsymbol{\delta}}_2^+\equiv
\frac{ \bar z_2^{-2}+{z}_1^{2}}{ \bar z_2^{-2}+\bar z_1^{2}}\left[\@I_2 
+\frac{({z}_1^{-2}+ \bar z_1^{2})( \bar z_1^{2}-{z}_1^{-2})
}{({z}_1^{-2}- \bar z_2^{-2})( \bar z_2^{2}+{z}_1^{-2})}
\frac{{\boldsymbol{\delta}}_1^-({\boldsymbol{\delta}}_1^{-})^\dagger}{\|{\boldsymbol{\delta}}_1^-\|^2}
\right]\tilde{\boldsymbol{\delta}}_2^+.
\end{gather}
\ese
Fig.~\ref{f:fs-cb} shows an example of an interaction of a fundamental soliton and a composite breather. Observe that, aside from a shift to its center, the fundamental soliton emerges unchanged. Fig.~\ref{f:fs-cb-asymp} shows the same interaction, with the asymptotics predicted by \eqref{fs-cb-shifts1} and \eqref{fs-cb-shifts2} subtracted.

Note that the rank of the norming constant is preserved in the interaction, so the composite breather remains a composite breather upon interacting with a rank-1 solution (be it a fundamental soliton or a fundamental breather). Indeed, one has 
%in which the composite (rank-2) breather would become rank-1 upon its interaction with the fundamental breather. Observe that
\begin{equation}
    \det\@C_{2}^{-}=\det\Big[\@c_{1}(z_{2},\boldsymbol\gamma_{1}^{-})\@C_{2}^{+}\@a_{1}(z_{2},\boldsymbol\delta_{1}^{-})\Big]=\det\@c_{1}(z_{2},\boldsymbol\gamma_{1}^{-})\det\@C_{2}^{+}\det\@a_{1}(z_{2},\boldsymbol\delta_{1}^{-})\,,
\end{equation}
proving that if $\@C_{2}^{+}$ is non-singular, so is $\@C_{2}^{-}$.

\section{Refactorization problem and Yang-Baxter maps}
In an interaction of $J$ fundamental breathers with velocities $v_{1}<v_{2}<\cdots<v_{J}$, the net change in the polarization of each soliton from $\tau\rightarrow-\infty$ to $\tau\rightarrow+\infty$ is independent of the order in which the intermediate interactions occur. As was established in \cite{APT06} in the special case of fundamental solitons, this result is tied to the fact that the polarization map \eqref{e:u+v-new} associated with each two-body interaction is a Yang-Baxter map. In this section, we reinterpret the two-body interactions in the discrete Manakov model in terms of Yang-Baxter maps by recasting the condition on the scattering coefficients that leads to the maps for the polarization vectors before and after the interaction as a refactorization problem. 

\subsection{Generalities}
We recall some general facts and notions from the theory of refactorization and Yang-Baxter maps, following e.g. \cite{V03,KP11}. 
Let $A (\@x, \lambda)$ be a given matrix-valued function
that depends on a point $\@x\in X$, with $X$ being some set (typically $\mathbb{C}\mathbb{P}^N$ or a Grassmannian in the context of soliton interactions), a (spectral) parameter $\lambda\in \Complex$, and possibly other (model-dependent) parameters. Consider the refactorization problem 
\begin{equation}
\label{e:YBf}
A(\@y^+,\lambda)A(\@x^+, \lambda) = A(\@x^-, \lambda)A(\@y^-,\lambda) \qquad \forall \lambda\in\Complex\,.
\end{equation}
If this uniquely defines $\@x^{+},\@y^{+}$ for each $\@x^{-},\@y^{-}$, then it gives rise to a map\footnote{With the simplified notations used here, it might look like this unique solution must be the trivial permutation solution $(\@x^+,\@y^+)=(\@y^-,\@x^-)$ but we will see that this is not the case in practice because of the other parameters involved in the model.}
$R:X\times X\to X\times X$, $(\@x^-
, \@y^-) \mapsto (\@x^+, \@y^+)$. If in addition the equation 
\begin{equation}
\label{triple_condition}
A(\@z,\lambda)A(\@y,\lambda)A(\@x, \lambda) = A(\@c, \lambda)A(\@b, \lambda)A(\@a,\lambda) \qquad \forall \lambda\in\Complex\,,
\end{equation}
implies $(\@x,\@y,\@z)=(\@a,\@b,\@c)$ then the map $R$ satisfies the Yang-Baxter equation on $X\times X\times X$, namely
$$R_{12}R_{13}R_{23}=R_{23}R_{13}R_{12}\,.$$
An important embodiment of these ideas arises in the context of loop groups in which the role of $A (\@x, \lambda)$ is played by the so-called simple element 
\begin{equation}
	\label{rational}
g_{\alpha_1,\alpha_2,\Pi}(\lda)=	\@I+\left(\frac{\lambda-\alpha_1}{\lambda-\alpha_2}-1\right)\Pi\,,
\end{equation}
where $\@I$ is the identity matrix of appropriate size, $\alpha_1,\alpha_2\in\CC$, $\alpha_1\neq \alpha_2$ and $\Pi^2=\Pi$ is a projector in some complex vector space. Note that its inverse is
\begin{equation}
	\label{inverse}
	g_{\alpha_1,\alpha_2,\Pi}(\lda)^{-1}=	\@I+\left(\frac{\lambda-\alpha_2}{\lambda-\alpha_1}-1\right)\Pi=g_{\alpha_2,\alpha_1,\Pi}(\lda)\,.
\end{equation}
The following results can be found in \cite{TU00}, see also \cite{L22,LC24}.
\begin{proposition}
Let $g_{\alpha_1,\alpha_2,\Pi_1}(\lda)$ and $g_{\beta_1,\beta_2,\Pi_2}(\lda)$ be two simple elements. If
\begin{equation}
	\label{phi}
	\phi=(\alpha_2-\beta_1)\@I+(\alpha_1-\alpha_2)\Pi_1+(\beta_1-\beta_2)\Pi_2
\end{equation}
is invertible, then	
\begin{equation}
\label{refactorization}	g_{\alpha_1,\alpha_2,\Pi_1}(\lda)g_{\beta_1,\beta_2,\Pi_2}(\lda)=g_{\beta_1,\beta_2,P_2}(\lda)g_{\alpha_1,\alpha_2,P_1}(\lda)
\end{equation}
if and only if
\begin{equation}
	\label{relation_projectors}
	P_i=\phi \Pi_i \phi^{-1}\,,~~i=1,2\,.
\end{equation}
\end{proposition}
For our purposes, the reduced case, whereby $\Pi$ is a Hermitian projector $\Pi^\dagger=\Pi$ and $\alpha_2=\alpha_1^*$, will be relevant. This is equivalent to the symmetry 
$$	g_{\alpha_1,\alpha_2,\Pi}(\lda)^{-1}=	g_{\alpha_1,\alpha_2,\Pi}(\lda^*)^\dagger\,,$$
and then it is enough to denote the simple element by $g_{\alpha_1,\Pi}(\lda)$. In this reduced case, \cite{TU00} shows that if $\alpha_1\neq \alpha_2$ and $\alpha_1\neq \alpha_2^*$ then $\phi$ is always invertible so that the refactorization problem 
\begin{equation}
\label{refactorization_red}	
g_{\alpha_1,\Pi_1}(\lda)g_{\alpha_2,\Pi_2}(\lda)=g_{\alpha_2,P_2}(\lda)g_{\alpha_1,P_1}(\lda)\,,
\end{equation}
is equivalent to the relation \eqref{relation_projectors} between the projectors, and that $P_1$, $P_2$ are also Hermitian projectors. Finally, condition \eqref{triple_condition} holds for such simple elements, see e.g. \cite{L22}. Thus, \eqref{refactorization_red} yields a parametric Yang-Baxter map $R(\alpha_1,\alpha_2): (\Pi_1,\Pi_2)\mapsto (P_1,P_2)$ given explicitly by 
\begin{equation}
\label{YB_map_proj}
    P_i=\phi \Pi_i \phi^{-1}\,,~~i=1,2\,,~~\phi= (\alpha_1^*-\alpha_2)\@I+(\alpha_1-\alpha_1^*)\Pi_1+(\alpha_2-\alpha_2^*)\Pi_2\,.
\end{equation}

\subsection{Application to interactions in the integrable discrete Manakov model}

The point is that the two-body interactions in the discrete Manakov model can be cast into the (reduced) refactorization problem \eqref{refactorization_red} where the role of the simple elements is played by the scattering coefficients in \eqref{e:CB_transm} and \eqref{e:FB_transm}. Given that for composite breathers the transmission coefficients \eqref{e:CB_transm} are proportional to the identity, the refactorization is trivial and no interesting Yang-Baxter map arises. Thus we focus on the fundamental breather coefficients (the fundamental soliton being a special case) and restrict our attention to \eqref{c} since the structure of \eqref{cbar} is similar. Specifically, consider the coefficients $\@c_{j}(z,\boldsymbol{\gamma}_{j})$ in \eqref{FB_coef} and the refactorization
\begin{equation}
\label{refac_c_j}
    \@c_{2}(z,\boldsymbol{\gamma}_{2}^{-})\@c_{1}(z,\boldsymbol{\gamma}_{1}^{-})=\@c_{1}(z,\boldsymbol{\gamma}_{1}^{+})\@c_{2}(z,\boldsymbol{\gamma}_{2}^{+})\,.
\end{equation}
We claim this is a special case of \eqref{refactorization_red}. First note that the factor $\alpha_{j}(z)$ in
$$\@c_{j}(z,\boldsymbol{\gamma}_{j})=\alpha_{j}(z)\bigg[\@I+\beta_{j}(z)\frac{\boldsymbol{\gamma}_{j}\boldsymbol{\gamma}_{j}^{\dagger}}{\boldsymbol{\gamma}_{j}^{\dagger}\boldsymbol{\gamma}_{j}}\bigg]\,,$$
plays no role since the product $\alpha_{1}(z)\alpha_{2}(z)$ appears on both sides of \eqref{refac_c_j}. Second, observe that
\begin{align}
	\beta_j(z)&=\frac{(z^2-z_j^2)(z^2+z_j^{-2})}{(z^2-\bar{z}_j^2)(z^2+\bar{z}_j^{-2})}-1 \nonumber\\
	&=\frac{z^2-z^{-2}-(z_j^2-z_j^{-2})}{z^2-z^{-2}-(\bar{z}_j^2-\bar{z}_j^{-2})}-1 \nonumber\\
	&=\frac{\sin 2\mu-\sin 2 \mu_j}{\sin 2 \mu -\sin 2 \mu_j^*}-1~~(z\to e^{i\mu}) \label{subs1}\\
	&=\frac{\lda-\lda_j}{\lda-\lda_j^*}-1\,,~~(\sin 2 \mu\to \lda ) \label{subs2}\,.
\end{align}
Third, the projectors here are rank one Hermitian projectors in $\Complex^2$ and the correspondence with \eqref{YB_map_proj} is $\alpha_j\to \lambda_j$, $\Pi_j\to \Pi_j^-=\frac{\boldsymbol{\gamma}^-_{j}\boldsymbol{\gamma}_{j}^{-\dagger}}{\boldsymbol{\gamma}_{j}^{-\dagger}\boldsymbol{\gamma}^-_{j}}$, $P_j\to \Pi_j^+=\frac{\boldsymbol{\gamma}^+_{j}\boldsymbol{\gamma}_{j}^{+\dagger}}{\boldsymbol{\gamma}_{j}^{+\dagger}\boldsymbol{\gamma}_{j}}$, $j=1,2$.
Therefore, we have a parametric Yang-Baxter map on rank one Hermitian projectors of {\it trigonometric type}, as seen from the reparametrization \eqref{subs1}. Strictly from the point of view of the YB maps, there is no essential new effect on the polarization shifts from this trigonometric map as compared for instance to the rational. Indeed, the last reparametrization \eqref{subs2} shows that in terms of $\lambda$, the YB map takes the same form as in the rational case. However, physically speaking, one has to remember that the trigonometric nature of the map is tied into the similar trigonometric nature of the Lax matrices in \eqref{e:Lax}. In turn, this dictates how the physical properties of the solitons such as their velocity and amplitudes are related to the spectral parameters (see e.g. \eqref{e:1FS}-\eqref{e:soliton_params}) and therefore how they influence the polarization shifts. Given the same velocity say of a soliton in the rational case and the trigonometric case, the corresponding effect on the polarization shifts when interacting with another soliton will be different in the two cases.

A rank one Hermitian projector is in one-to-one correspondence with a nonzero vector modulo its norm, i.e. with an element of $\mathbb{C}\mathbb{P}^1$, and we can deduce a trigonometric Yang-Baxter map on $\mathbb{C}\mathbb{P}^1$ from the map we have just described:
\begin{eqnarray}
	\label{map_proj}
	\Pi_j^+=\phi \Pi_j^- \phi^{-1}\,,~~\phi= (\lambda_1^*-\lambda_2)\@I+(\lambda_1-\lambda_1^*)\Pi_1+(\lambda_2-\lambda_2^*)\Pi_2\,,
\end{eqnarray}
where we recall that $\lambda_j=\frac{z_j^2-z_j^{-2}}{2i}$ in the original parametrization. 
It remains to derive the desired map between $\boldsymbol{\gamma}_j^+$ and $\boldsymbol{\gamma}_j^-$ (up to normalization), using $\Pi_j^\pm=\frac{\boldsymbol{\gamma}_j^\pm\boldsymbol{\gamma}_j^{\pm\dagger}}{\boldsymbol{\gamma}_j^{\pm\dagger}\boldsymbol{\gamma}_j^{\pm}}$. For convenience, write $\alpha_1= r_1+is_1$, $\alpha_2= r_2+is_2$ so that
\begin{equation}
	\phi=(r_1-r_2)\@I+is_1(2\Pi^-_1-\@I)+is_2(2\Pi^-_2-\@I)\equiv r\@I+i(s_1\sigma_1+s_2\sigma_2)\,.
\end{equation}
Note that $\sigma_j$ are Hermitian involutions. Also,  $\phi^\dagger\phi=\phi\phi^\dagger=(r^2+s_1^2+s_2^2)\1+s_1s_2(\sigma_1\sigma_2+\sigma_2\sigma_1)$
and, as a result, we have
$$\phi^\dagger\phi \,\sigma_j=\sigma_j\,\phi^\dagger\phi\quad \Rightarrow \phi^\dagger\phi \quad \Pi^-_j=\Pi^-_j\,\phi^\dagger\phi\,.$$
Thus, on the one hand
\begin{equation}
\Pi^-_j\,	\phi^\dagger\phi \,\Pi^-_j=\phi^\dagger\phi \,\Pi^-_j\,,
\end{equation}
and on the other hand, direct calculation gives
\begin{align}
	\Pi^-_j\,	\phi^\dagger\phi \,\Pi^-_j&=(r^2+s_1^2+s_2^2)\Pi^-_j+s_1s_2\Pi^-_j(\sigma_1\sigma_2+\sigma_2\sigma_1)\Pi^-_j\nonumber\\
	&=\left(r^2+(s_1-s_2)^2+4s_1s_2\frac{|\gamma_1^{-\dagger}\gamma_2^-|^2}{||\gamma_1^-||^2||\gamma_2^-||^2} \right)\Pi^-_j\nonumber\\
	&\equiv \Delta^2 \,\Pi^-_j\,.
\end{align}
	Essentially the same calculation gives
	\begin{equation}
||\phi\boldsymbol{\gamma}_j^-||^2=		\boldsymbol{\gamma}_j^{-\dagger} \,	\phi^\dagger\phi \,\boldsymbol{\gamma}^-_j=\Delta^2 \,||\boldsymbol{\gamma}_j^-||^2\,.
	\end{equation}
With this, \eqref{map_proj} yields
\begin{eqnarray}	\frac{\boldsymbol{\gamma}_j^+\boldsymbol{\gamma}_j^{+\dagger}}{||\boldsymbol{\gamma}_j^+||^2}=\Pi_j^+=\phi \Pi_j^- \phi^{-1}=\frac{1}{\Delta^2}\phi \Pi_j^- \phi^{\dagger}=\frac{\phi \boldsymbol{\gamma}_j^-(\phi \boldsymbol{\gamma}_j^-)^\dagger}{||\phi \boldsymbol{\gamma}_j^-||^2}\,.
\end{eqnarray}
Thus, up to a normalization constant $\mu_j$ we have the map
\begin{equation}
	\boldsymbol{\gamma}_j^+=\mu_j\phi \gamma_j^-\,.
\end{equation}
More explicitly, using the expression in \eqref{map_proj} for $\phi$ we have, forgetting about the normalizations,
\begin{align}
	\boldsymbol{\gamma}_1^+=\left(\@I+\frac{\lda_2-\lda_2^*}{\lda_1-\lda_2}\Pi_2^-\right)\boldsymbol{\gamma}_1^- , \qquad
	\boldsymbol{\gamma}_2^+=\left(\@I+\frac{\lda_1-\lda_1^*}{\lda_1^*-\lda_2^*}\Pi_1^-\right)\boldsymbol{\gamma}_2^-,  \label{map1&2}
\end{align}
which we can interpret as a map between elements $[\boldsymbol{\gamma}_j^\pm]$ in $\mathbb{C}\mathbb{P}^1$ given in terms of representatives $\boldsymbol{\gamma}_j^\pm$ in $\Complex^2$, if one is not interested in normalizations. Of course, physically it is important to determine the normalizations precisely, and the corresponding maps are then the ones given in \eqref{e:u+v-new1}-\eqref{e:u+v-new2}.

\subsection{Common structures and differences across three integrable models}
It is important to point out that the vector NLS, ccSPE and discrete Manakov all fall into the above scheme, in the sense that looking at the refactorization of the scattering coefficients amounts to \eqref{refactorization_red}, in the special case of certain reductions and with appropriate reparametrization of the spectral parameters. In turn, each version of \eqref{refactorization_red} adapted to the model at hand plays a crucial role in the description of soliton interactions in the model. Before proceeding with the description of the common features of the three models, we first discuss an important difference. The solitons in the (focusing) vector NLS (on the full line with decaying boundary conditions) only correspond to rank one projectors in the refactorization problem \eqref{refactorization_red}, while ccSPE and discrete Manakov both also support fundamental and composite breathers (fundamental solitons are just a special case of fundamental breathers). Composite breathers do not lead to interesting maps on the polarizations so we do not dwell on them in this discussion.

Thus, the common structure between the three models is the following type of transmission coefficients
$$\@c_j(k,\boldsymbol\gamma)=\alpha_j(k)\left(\@I+\beta_j(k)\Pi\right)\,,$$
where $\Pi=\frac{\boldsymbol{\gamma}\boldsymbol{\gamma}^\dagger}{\boldsymbol{\gamma}^\dagger\boldsymbol{\gamma}}$ is a rank one Hermitian projector and 
\begin{align}
    & \beta_j(k)=\frac{k_j^*-k_j}{k-k_j^*}~~\text{(vector NLS)}\,,\\
    &\beta_j(k)=\frac{k^2}{k_j^2}\frac{k_j^{*2}-k_j^2}{k^2-k_j^{*2}}~~\text{(ccSPE)}\,,\\
    &\beta_j(k)=\frac{(\bar{k}_j^2-k_j^2)(\bar{k}_j^2-k_j^{-2})}{(k^2-\bar{k}_j^2)(k^{-2}+\bar{k}_j^2)}~~ \text{(discrete Manakov)}\,.
\end{align}
In all three cases, the refactorization property of the scattering coefficients
\begin{equation}
\label{refactorization_c}	
\@c_1(k,\boldsymbol\gamma_1^+)\@c_2(k,\boldsymbol\gamma_2^+)=\@c_2(k,\boldsymbol\gamma_2^-)\@c_1(k,\boldsymbol\gamma_1^-)\,,
\end{equation}
is equivalent to \eqref{refactorization_red}. We explained this in \eqref{subs1}-\eqref{subs2} for the discrete Manakov model. For the vector NLS and ccSPE, it suffices to note that
\begin{align}
    & \beta_j(k)=\frac{\lda-\lda_j}{\lda-\lda_j^*}-1\,,~~k\to\lda~~\text{(vector NLS)}\,,\\
    &	\beta_j(k)=\frac{\frac{1}{k^2}-\frac{1}{k_j^2}}{\frac{1}{k^2}-\frac{1}{k_j^{*2}}}-1= \frac{\lda-\lda_j}{\lda-\lda_j^*}-1\,,~~1/k^2\to\lda ~~\text{(ccSPE)}\,.
\end{align}
In the terminology that originated in \cite{BD82} and is widespread in (quantum) integrable systems, the vector NLS provides Yang-Baxter maps of rational type, while we found in the present work that the discrete Manakov model provides Yang-Baxter maps of trigonometric type. 

\section{Concluding remarks} 

In this work, we presented a complete characterization of the various soliton and breather interactions in the integrable discrete Manakov model. This was done using a generalization of the Manakov method (which essentially expresses the change in the polarization of the soliton/breather after the interaction via the transmission coefficients associated to the interacting soliton/breather) to include fundamental and composite breathers. 
As is well-known, interactions in multicomponent integrable PDEs (and, as shown here, in their integrable discretizations) are intimately related to set-theoretical solutions of the Yang-Baxter equation with spectral parameters, or parametric Yang-Baxter maps.
In particular, the explicit formulas for the transmission coefficients that characterize fundamental solitons  and fundamental breathers allowed us to interpret the interactions in terms of a refactorization problem, and derive the associated Yang-Baxter maps describing the effect of interactions on the soliton polarizations. As an essential feature of the integrable discrete Manakov model, a novel Yang–Baxter map of trigonometric type was derived. In that respect, we expect these maps to possess nice Poisson and symplectic properties arising from the underlying Hamiltonian structure of the discrete Manakov model. The latter would be the natural vector generalization of the one presented e.g. in \cite{CC19} which relied on the trigonometric classical $r$-matrix in the sense of the Belavin-Drinfeld classification \cite{BD82}. Such nice properties were studied in detail in \cite{L22,LC24} in relation to the rational case. The investigation of these properties in the trigonometric case is beyond the scope of this paper, but we expect it will be the subject of future work.

Another natural follow-up question concerns the type of set-theoretical solutions of the reflection equation, or reflection maps \cite{CZ14,CCZ13}, that could be constructed from the trigonometric Yang-Baxter map found here. This is particularly relevant as a fruitful method to construct such reflection maps is to pose the integrable model on the half-line (or positive integers) and study the corresponding integrable boundary conditions. Since the Ablowitz-Ladik with such boundary conditions has been thoroughly studied \cite{BH10,BB15,CC19}, we expect that an extension to the discrete Manakov model would yield reflection maps of trigonometric type. In turn, this is of relevance to the study of the set-theoretical Yang-Baxter and reflection equations in the context of (skew)-braces and other related algebraic structures. Recently, the importance of developing parametric versions of these structures has emerged in \cite{D24,DMS24}. This is left for future work.  

Finally, we remark that the 2-component vector system \eqref{e:idMq} has an integrable $N$-component generalization (see for example \cite{TWU99}), and a study similar to that of the present paper could be applied to this generic case. The norming constants would be $N\times N$ matrices, so one could consider the interactions of solitons associated with $N$ possible ranks. In general, this would produce YB maps relating to Grassmannians and the variety of possible interactions between solitons corresponding to different ranks of projectors will increase substantially. 

\section*{Acknowledgements}

BP and NJO gratefully acknowledge partial support for this work from the NSF, under grant DMS-2406626.
The authors also acknowledge the Isaac Newton Institute for Mathematical Sciences, Cambridge, UK, for support and hospitality during the satellite program ``Emergent phenomena in nonlinear dispersive waves'' (supported by the EPSRC grant EP/R014604/1), where work on this paper was carried out. Finally, we would also like to thank Dr Carlos Dibaya, who participated in some of the initial proceedings of this study.

\nopagebreak 
\appendix

\section{Derivation of the transmission coefficients}

\paragraph{Fundamental breathers.} In the case of the single fundamental breather solution, where $\@C_{1}=\boldsymbol\gamma_{1}\boldsymbol\delta_{1}^{\dagger}$ and $\hat{\@C}_{1}\@C_{1}=\bar{\@C}_{1}\tilde{\@C}_{1}=0$, the full solution of the linear systems \eqref{up_system} and \eqref{dn_system} with $J=1$ is given by:
\bse
\begin{gather}
    \bar{\@N}_n^{\prime\, \textrm{(up)}}(\bar z_{1}) = \frac{1}{1+\tilde{g}_{n}}\bigg[\@I_{2}+\tilde{g}_{n}\frac{\bar{z}_{1}^{2}+z_{1}^{-2}}{\bar{z}_{1}^{2}+\bar{z}_{1}^{-2}}\bigg(\@I_{2}-\frac{\boldsymbol\delta_{1}\boldsymbol\delta_{1}^{\dagger}}{||\boldsymbol\delta_{1}||^{2}}\bigg)\bigg]\,,\\
    \bar{\@N}_n^{\prime\, \textrm{(up)}}(i/z_{1}) = \frac{1}{1+\tilde{g}_{n}}\bigg[\@I_{2}+\tilde{g}_{n}\frac{{z}_{1}^{-2}+\bar z_{1}^{2}}{{z}_{1}^{-2}+{z}_{1}^{2}}\frac{\boldsymbol\delta_{1}\boldsymbol\delta_{1}^{\dagger}}{||\boldsymbol\delta_{1}||^{2}}\bigg]\,,
    \end{gather}
    \begin{gather}
    {\@N}_n^{\prime\, \textrm{(up)}}(z_{1}) = \frac{2z_{1}}{1+\tilde{g}_{n}}\bigg[\frac{\bar{z}_{1}^{2n+2}}{z_{1}^{2}-\bar{z}_{1}^{2}}\boldsymbol\delta_{1}\boldsymbol\gamma_{1}^{\dagger}+(-1)^{n}\frac{z_{1}^{-2n-2}}{z_{1}^{2}+z_{1}^{-2}}\big(\boldsymbol\delta_{1}^{\dagger}\boldsymbol\gamma_{1}\@I_{2}-\boldsymbol\gamma_{1}\boldsymbol\delta_{1}^{\dagger}\big)\bigg]\,,\\
    {\@N}_n^{\prime\, \textrm{(up)}}(i/\bar z_{1}) = \frac{-2i\bar z_{1}^{-1}}{1+\tilde{g}_{n}}\bigg[\frac{\bar{z}_{1}^{2n+2}}{\bar z_{1}^{-2}+\bar{z}_{1}^{2}}\boldsymbol\delta_{1}\boldsymbol\gamma_{1}^{\dagger}+(-1)^{n}\frac{z_{1}^{-2n-2}}{\bar z_{1}^{-2}-z_{1}^{-2}}\big(\boldsymbol\delta_{1}^{\dagger}\boldsymbol\gamma_{1}\@I_{2}-\boldsymbol\gamma_{1}\boldsymbol\delta_{1}^{\dagger}\big)\bigg]\,, 
        \end{gather}
    \begin{gather}
    \bar{\@N}_n^{\prime\, \textrm{(dn)}}(\bar z_{1}) = \frac{2\bar z_{1}}{1+\tilde{g}_{n+1}}\bigg[\frac{{z}_{1}^{-2n}}{\bar z_{1}^{2}-{z}_{1}^{2}}\boldsymbol\gamma_{1}\boldsymbol\delta_{1}^{\dagger}+(-1)^{n+1}\frac{\bar z_{1}^{2n}}{\bar z_{1}^{2}+\bar z_{1}^{-2}}\big(\boldsymbol\gamma_{1}^{\dagger}\boldsymbol\delta_{1}\@I_{2}-\boldsymbol\delta_{1}\boldsymbol\gamma_{1}^{\dagger}\big)\bigg]\,,\\
    \bar{\@N}_n^{\prime\, \textrm{(dn)}}(i/z_{1}) = \frac{-2iz_{1}^{-1}}{1+\tilde{g}_{n+1}}\bigg[\frac{{z}_{1}^{-2n}}{z_{1}^{-2}+{z}_{1}^{2}}\boldsymbol\gamma_{1}\boldsymbol\delta_{1}^{\dagger}+(-1)^{n+1}\frac{\bar z_{1}^{2n}}{z_{1}^{-2}-\bar z_{1}^{-2}}\big(\boldsymbol\gamma_{1}^{\dagger}\boldsymbol\delta_{1}\@I_{2}-\boldsymbol\delta_{1}\boldsymbol\gamma_{1}^{\dagger}\big)\bigg]\,,
    \end{gather}
    \begin{gather}
    {\@N}_n^{\prime\, \textrm{(dn)}}(z_{1}) = \frac{1}{1+\tilde{g}_{n+1}}\bigg[\@I_{2}+\tilde{g}_{n+1}\frac{{z}_{1}^{2}+\bar z_{1}^{-2}}{{z}_{1}^{2}+{z}_{1}^{-2}}\bigg(\@I_{2}-\frac{\boldsymbol\gamma_{1}\boldsymbol\gamma_{1}^{\dagger}}{||\boldsymbol\gamma_{1}||^{2}}\bigg)\bigg]\,,\\
    {\@N}_n^{\prime\, \textrm{(dn)}}(i/\bar z_{1}) = \frac{1}{1+\tilde{g}_{n+1}}\bigg[\@I_{2}+\tilde{g}_{n+1}\frac{\bar{z}_{1}^{-2}+z_{1}^{2}}{\bar{z}_{1}^{-2}+\bar{z}_{1}^{2}}\frac{\boldsymbol\gamma_{1}\boldsymbol\gamma_{1}^{\dagger}}{||\boldsymbol\gamma_{1}||^{2}}\bigg]\,,
\end{gather}
\ese
where 
\begin{equation}
    \tilde{g}_{n}=4\frac{z_{1}^{-2n+2}\bar{z}_{1}^{2n+2}}{(z_{1}^{2}-\bar{z}_{1}^{2})^{2}}||\boldsymbol\gamma_{1}||^{2}||\boldsymbol\delta_{1}||^{2}\,.
\end{equation}
With this, from \eqref{Nbar_RHP} one can compute the explicit form of the eigenfunctions for all $z$:
\bse
\begin{eqnarray}
    \bar{\@N}_n^{\prime\, \textrm{(up)}}(z) &=& \frac{1}{1+\tilde{g}_{n}}\bigg[\bigg(1+\tilde{g}_{n}\frac{z^{2}+z_{1}^{-2}}{z^{2}+\bar{z}_{1}^{-2}}\bigg)\@I_{2}+\tilde{g}_{n}\frac{(\bar{z}_{1}^{2}+z_{1}^{-2})(\bar{z}_{1}^{-2}-z_{1}^{-2})}{(z_{1}^{-2}-z^{-2})(z^{2}+\bar{z}_{1}^{-2})}\frac{\boldsymbol\delta_{1}\boldsymbol\delta_{1}^{\dagger}}{||\boldsymbol\delta_{1}||^{2}}\bigg]\,,\\
    {\@N}_n^{\prime\, \textrm{(up)}}(z) &=& \frac{2}{1+\tilde{g}_{n}}\bigg[\frac{\bar{z}_{1}^{2n+2}z}{z^{2}-\bar{z}_{1}^{2}}\boldsymbol\delta_{1}\boldsymbol\gamma_{1}^{\dagger}+(-1)^{n}\frac{z_{1}^{-2n-2}z}{z^{2}+{z}_{1}^{-2}}\big(\boldsymbol\delta_{1}^{\dagger}\boldsymbol\gamma_{1}\@I_{2}-\boldsymbol\gamma_{1}\boldsymbol\delta_{1}^{\dagger}\big)\bigg]\,,\\
    \bar{\@N}_n^{\prime\, \textrm{(dn)}}(z) &=&\frac{2}{1+\tilde{g}_{n+1}}\bigg[\frac{z_{1}^{-2n}z}{z^{2}-z_{1}^{2}}\boldsymbol\gamma_{1}\boldsymbol\delta_{1}^{\dagger}+(-1)^{n+1}\frac{\bar{z}_{1}^{2n}z}{z^{2}+\bar{z}_{1}^{-2}}\big(\boldsymbol\gamma_{1}^{\dagger}\boldsymbol\delta_{1} \@I_{2}-\boldsymbol\delta_{1}\boldsymbol\gamma_{1}^{\dagger}\big)\bigg]\,,\\
    {\@N}_n^{\prime\, \textrm{(dn)}}(z) &=& \frac{1}{1+\tilde{g}_{n+1}}\bigg[\bigg(1+\tilde{g}_{n}\frac{\bar{z}_{1}^{2}+z^{-2}}{z_{1}^{2}+z^{-2}}\bigg)\@I_{2}+\tilde{g}_{n}\frac{(\bar{z}_{1}^{2}-{z}_{1}^{2})(\bar{z}_{1}^{2}+z_{1}^{-2})}{(z^{2}-\bar{z}_{1}^{2})(z_{1}^{2}+z^{-2})}\frac{\boldsymbol\gamma_{1}\boldsymbol\gamma_{1}^{\dagger}}{||\boldsymbol\gamma_{1}||^{2}}\bigg]\,.
\end{eqnarray}
\ese
Additionally, from \eqref{e:Deltan}, it can found that
\begin{equation}
    \Delta_{n}=\frac{1+\tilde{g}_n}{1+\tilde g_{n+1}}\,,
\end{equation}
so that in light of \eqref{e:Nprimes}, the original eigenfunctions are
\bse
\begin{eqnarray}
    &&\bar{\@N}_n^{ \textrm{(up)}}(z)=\bar{\@N}_n^{\prime\, \textrm{(up)}}(z),\;\;\;{\@N}_n^{\textrm{(up)}}(z)={\@N}_n^{\prime\, \textrm{(up)}}(z)\,,\\
        \bar{\@N}_n^{\textrm{(dn)}}(z) &=&\frac{2}{1+\tilde{g}_{n}}\bigg[\frac{z_{1}^{-2n}z}{z^{2}-z_{1}^{2}}\boldsymbol\gamma_{1}\boldsymbol\delta_{1}^{\dagger}+(-1)^{n+1}\frac{\bar{z}_{1}^{2n}z}{z^{2}+\bar{z}_{1}^{-2}}\big(\boldsymbol\gamma_{1}^{\dagger}\boldsymbol\delta_{1} \@I_{2}-\boldsymbol\delta_{1}\boldsymbol\gamma_{1}^{\dagger}\big)\bigg]\,,\\
    {\@N}_n^{\textrm{(dn)}}(z) &=& \frac{1}{1+\tilde{g}_{n}}\bigg[\bigg(1+\tilde{g}_{n}\frac{\bar{z}_{1}^{2}+z^{-2}}{z_{1}^{2}+z^{-2}}\bigg)\@I_{2}+\tilde{g}_{n}\frac{(\bar{z}_{1}^{2}-{z}_{1}^{2})(\bar{z}_{1}^{2}+z_{1}^{-2})}{(z^{2}-\bar{z}_{1}^{2})(z_{1}^{2}+z^{-2})}\frac{\boldsymbol\gamma_{1}\boldsymbol\gamma_{1}^{\dagger}}{||\boldsymbol\gamma_{1}||^{2}}\bigg]\,.
\end{eqnarray}
\ese
To compute the relevant transmission coefficients $\bar{\@c}(z)$ and $\@c(z)$, one simply needs to take the limit as $n\rightarrow-\infty$ of $\bar{\@N}_n^{ \textrm{(up)}}(z)$ and ${\@N}_n^{\textrm{(dn)}}(z)$, respectively. With this in mind, note that $\tilde{g}_{n}$ grows exponentially in this limit. In particular, if $z_{1}=\exp(a_{1}+ib_{1})$ then $\tilde{g}_{n}=re^{-4a_{1}n}$ (recall that $a_{1}$ is assumed to be positive) with $r$ independent of $n$. The formulas for the transmission coefficients given in \eqref{e:FB_transm} then clearly follow from
\begin{gather}
 %\hspace*{-3mm}   
 \bar{\@c}(z)=\lim_{n\rightarrow-\infty}\frac{1}{1+re^{-4a_{1}n}}\bigg[\bigg(1+re^{-4a_{1}n}\frac{z^{2}+z_{1}^{-2}}{z^{2}+\bar{z}_{1}^{-2}}\bigg)\@I_{2}+re^{-4a_{1}n}\frac{(\bar{z}_{1}^{2}+z_{1}^{-2})(\bar{z}_{1}^{-2}-z_{1}^{-2})}{(z_{1}^{-2}-z^{-2})(z^{2}+\bar{z}_{1}^{-2})}\frac{\boldsymbol\delta_{1}\boldsymbol\delta_{1}^{\dagger}}{||\boldsymbol\delta_{1}||^{2}}\bigg]\,,\\
    \@c(z)=\lim_{n\rightarrow-\infty}\frac{1}{1+re^{-4a_{1}n}}\bigg[\bigg(1+re^{-4a_{1}n}\frac{\bar{z}_{1}^{2}+z^{-2}}{z_{1}^{2}+z^{-2}}\bigg)\@I_{2}+re^{-4a_{1}n}\frac{(\bar{z}_{1}^{2}-{z}_{1}^{2})(\bar{z}_{1}^{2}+z_{1}^{-2})}{(z^{2}-\bar{z}_{1}^{2})(z_{1}^{2}+z^{-2})}\frac{\boldsymbol\gamma_{1}\boldsymbol\gamma_{1}^{\dagger}}{||\boldsymbol\gamma_{1}||^{2}}\bigg]\,.
\end{gather}

\paragraph{Composite breathers.} The transmission coefficients associated with a composite breather solution can be obtained using the same procedure; though in this case, since $\@C_{1}$ is generic, the expressions for the eigenfunctions are not as concise as in the fundamental breather case. For instance, upon solving the linear system for $\bar{\@N}_n^{\prime\, \textrm{(up)}}(\bar z_{1})$ and $\bar{\@N}_n^{\prime\, \textrm{(up)}}(i/z_{1})$, one finds
\bse
\begin{gather}
    \bar{\@N}_n^{\prime\, \textrm{(up)}}(\bar z_{1})=\frac{1}{1+g_{n}}\bigg[\@I_{2}-4(-1)^{n}\frac{z_{1}^{-4n}(1+z_{1}^{2}\bar{z}_{1}^{2})}{(z_{1}^{-2}+z_{1}^{2})^{2}(z_{1}^{2}-\bar{z}_{1}^{2})}\hat{\@C}_{1}\@C_{1}-4\frac{z_{1}^{-2n+2}\bar{z}_{1}^{2n+2}(1+z_{1}^{2}\bar{z}_{1}^{2})}{(z_{1}^{2}-\bar{z}_{1}^{2})^{2}(\bar{z}_{1}^{-2}+\bar{z}_{1}^{2})}\hat{\@C}_{1}\tilde{\@C}_{1}\bigg],\,\\
    \bar{\@N}_n^{\prime\, \textrm{(up)}}(i/z_{1})=\frac{1}{1+g_{n}}\bigg[\@I_{2}+4\frac{z_{1}^{-2n}\bar{z}_{1}^{2n}(1+z_{1}^{2}\bar{z}_{1}^{2})}{(z_{1}^{-2}+z_{1}^{2})(z_{1}^{2}-\bar{z}_{1}^{2})^{2}}\bar{\@C}_{1}\@C_{1}+4(-1)^{n}\frac{\bar{z}_{1}^{4n}(1+z_{1}^{2}\bar{z}_{1}^{2})}{({z}_{1}^{2}-\bar z_{1}^{2})(\bar{z}_{1}^{-2}+\bar{z}_{1}^{2})^{2}}\bar{\@C}_{1}\tilde{\@C}_{1}\bigg]\,.
\end{gather}
\ese
Here, $g_{n}$ is the same as in the composite breather solution, defined in \eqref{g_composite}. The above can be used to construct a lengthy expression for $\bar{\@N}_n^{ \textrm{(up)}}(z)$, and the limit as $n\rightarrow-\infty$ can be computed to find $\bar{\@c}(z)$. In doing so, one finds that when $\@C_{1}$ is rank-2, the fastest growing terms in the numerator and denominator of $\bar{\@N}_n^{ \textrm{(up)}}(z)$ are at $\mathcal{O}(e^{-8a_{1}n})$, and hence dominate the $\mathcal{O}(e^{-4a_{1}n})$ terms that contributed to the limit in the rank-1 case.

\section{Long-time asymptotics for the 2-fundamental breather solution}

As mentioned in Section 4, the two long-time limits of the 2-soliton solution that are straightforward to compute are the $\tau\rightarrow-\infty$ limit in the reference frame of soliton 1 (the slower soliton) and the $\tau\rightarrow+\infty$ limit in the reference frame of soliton 2 (the faster soliton). Here we show the derivation of the former limit, and the latter follows from a similar calculation. 

The 2-fundamental breather solution can be obtained from \eqref{potential} after solving the linear system \eqref{up_system_1}-\eqref{up_system_4}, all with $J=2$. Upon substituting \eqref{up_system_3}-\eqref{up_system_4} into \eqref{up_system_1}-\eqref{up_system_2}, all terms involving $\hat{\@C}_{j}\@C_{j}$ or $\bar{\@C}_{j}\tilde{\@C}_{j}$, which are zero in the rank-1 case, can be dropped. Additionally, in the limit $\tau\rightarrow-\infty$ with $\zeta_{1}$ (as defined in \eqref{xi}) fixed, all terms involving $z_{2}^{-2n}\@C_{2}$, $\bar z_{2}^{2n}\bar{\@C}_{2}$, $z_{2}^{-2n}\hat{\@C}_{2}$, or $\bar z_{2}^{2n}\tilde{\@C}_{2}$ decay exponentially and can be neglected. From \eqref{Cs_xi}, using the fact that $\@C_{1}=\boldsymbol\gamma_{1}\boldsymbol\delta_{1}^{\dagger}$, the nonzero products of norming constants are
\begin{eqnarray}
        \bar{z}_{1}^{2n}z_{1}^{-2n}\bar{\@C}_{1}(\tau)\@C_{1}(\tau)&=&e^{-2\zeta_{1}}\bar{z}_{1}^{2}||\boldsymbol\gamma_{1}||^{2}||\boldsymbol\delta_{1}||^{2}\boldsymbol\Pi\,, \\
    {z}_{1}^{-2n}\bar z_{1}^{2n}\hat{\@C}_{1}(\tau)\tilde{\@C}_{1}(\tau)&=&-e^{-2\zeta_{1}}z_{1}^{-2}||\boldsymbol\gamma_{1}||^{2}||\boldsymbol\delta_{1}||^{2}(\@{I}_{2}-\boldsymbol\Pi)\,,
\end{eqnarray}
where 
\begin{equation}
\boldsymbol\Pi\equiv\frac{\boldsymbol\delta_{1}\boldsymbol\delta_{1}^{\dagger}}{||\boldsymbol\delta_{1}||^{2}},\;\;\;\boldsymbol\Pi^{2}=\boldsymbol\Pi\,.    
\end{equation}
With all of this in mind, one finds that after dropping all negligible terms, the desired asymptotic limit can be obtained from 
\begin{equation}
\label{asymp_potential}
\@Q_{n-1}(\tau)\sim-2\bar{z}_{2}^{-2}\@X\bar{\@C}_{1}e^{-\zeta_{1}+2ib_{1}n-2i\omega_{1}\tau}-2(-1)^{n-1}z_{1}^{2}\@Y\hat{\@C}_{1}e^{-\zeta_{1}-2ib_{1}n+2i\omega_{1}\tau}\,,
\end{equation}
where $\@X\equiv\bar{\@N}_n^{\prime\, \textrm{(up)}}(\bar z_{1})$ and $\@Y\equiv\bar{\@N}_n^{\prime\, \textrm{(up)}}(i/z_{1})$ satisfy the system
\begin{equation}
\label{asymp_system}
    \begin{cases}\@X(\@I_{2}+a\boldsymbol\Pi)+\@Yb(\@I_{2}-\boldsymbol\Pi)=\@I_{2}\\
    \@Xc\boldsymbol\Pi+\@Y\big(\@I_{2}+a(\@I_{2}-\boldsymbol\Pi)\big)=\@I_{2}\end{cases}\,,
\end{equation}
\begin{equation}
\label{a_def}
    a=4\frac{z_{1}^{2}\bar{z}_{1}^{2}}{(\bar{z}_{1}^{2}-z_{1}^{2})^{2}}e^{-2\zeta_{1}}||\boldsymbol\gamma_{1}||^{2}||\boldsymbol\delta_{1}||^{2},\;\;\;b=c^{*}=4\frac{\bar{z}_{1}^{-2}z_{1}^{-2}}{(\bar{z}_{1}^{2}+\bar{z}_{1}^{-2})(\bar{z}_{1}^{-2}-z_{1}^{-2})}e^{-2\zeta_{1}}||\boldsymbol\gamma_{1}||^{2}||\boldsymbol\delta_{1}||^{2}\,.
\end{equation}
Making use of the fact that
\begin{equation}
    (\@I_{2}+a\boldsymbol\Pi)^{-1}=\frac{1}{1+a}\Big(\@I_{2}+a(\@I_{2}-\boldsymbol\Pi)\Big)\,,
\end{equation}
the solution of the system \eqref{asymp_system} is
\bse
\begin{eqnarray}
    \@X&=&\@I_{2}-\frac{a}{1+a}\boldsymbol\Pi-\frac{b}{1+a}(\@I_{2}-\boldsymbol\Pi)\,,\\
    \@Y&=&\@I_{2}-\frac{c}{1+a}\boldsymbol\Pi-\frac{a}{1+a}(\@I_{2}-\boldsymbol\Pi)\,.
\end{eqnarray}
\ese
Considering \eqref{asymp_potential}, note that $\boldsymbol\Pi\bar{\@C}_{1}=\bar{\@C}_{1}$ and $\boldsymbol\Pi\hat{\@C}_{1}=\boldsymbol{0}$, so that 
\begin{equation}
    \@X\bar{\@C}_{1}=\frac{1}{1+a}\bar{\@C}_{1},\;\;\;\@Y\hat{\@C}_{1}=\frac{1}{1+a}\hat{\@C}_{1}\,.
\end{equation}
Substituting these into \eqref{asymp_potential} gives
\begin{equation}
    \@Q_{n-1}(\tau)\sim\frac{-2e^{-\zeta_{1}}}{1+a}\Big[\bar{z}_{1}^{-2}\bar{\@C}_{1}e^{2ib_{1}n-2i\omega_{1}\tau}+(-1)^{n-1}z_{1}^{2}\hat{\@C}_{1}e^{-2ib_{1}n+2i\omega_{1}\tau}\Big]\qquad \tau\to -\infty, \quad \text{fixed } \zeta_1\,,
\end{equation}
or in explicit vector form:
\begin{equation}
\label{asymp_fb}
    \begin{pmatrix}
Q_{n-1}^{(1)}(\tau) \\  Q_{n-1}^{(2)}(\tau)
\end{pmatrix}\sim\frac{-2e^{-\zeta_{1}}}{1+a}\Big[\mu_{1}^{*}\boldsymbol\gamma_{1}^{*}e^{2ib_{1}n-2i\omega_{1}\tau}+(-1)^{n}\kappa_{1}\boldsymbol\gamma_{1}^{\perp}e^{-2ib_{1}n+2i\omega_{1}\tau}\Big]\,.
\end{equation}
Upon plugging in the definition of $a$ from \eqref{a_def}, after simplification one can verify that \eqref{asymp_fb} is identical to the 1-fundamental breather solution in \eqref{fb}.

Although conceptually similar, computing the long-time asymptotics in the opposite limits is quite cumbersome because now the system has growing exponential terms, and at least 2 next-to-leading order terms have to be computed and retained at each step, since the system becomes degenerate. We give below some details on how the limit is computed for fixed $\zeta_2$ and $\tau \to -\infty$ (where, according to \eqref{e:zeta1vszeta2}, $e^{-\zeta_1}\to +\infty$) in the case in which both solitons are fundamental solitons. The basic idea is to expand the coefficients of the system \eqref{single_equ} with $J=2$, $\@C_j=\boldsymbol{\gamma}_j\boldsymbol{\delta}_j^\dagger$ for $j=1,2$ and $\boldsymbol{\delta}_j=(1, 0)^{T}$ in powers of $e^{-\zeta_1}$ (i.e. $\mathcal{O}(1), \mathcal{O}(e^{\zeta_1}), \dots$), iteratively solving the equations for one of the unknowns and back-substituting into the other equations, while keeping at least two non-zero terms at each step. This yields:
\bse
\label{e:LTA_2FB}
\begin{gather}
\label{lta_N1_1}
     \bar{\@N}^{\prime\, \mathrm{(up)}}_n(\bar{z}_1)\sim
    \begin{pmatrix}
    -e^{\zeta_1-\zeta_2}\displaystyle{\frac{l_4}{l_1}\frac{H_7}{H_1}}& \mathcal{O}(1) \\
     -e^{\zeta_1-\zeta_2}\displaystyle{\frac{l_{10}}{l_1}\frac{I_{13}}{I_6}} & \mathcal{O}(1)
    \end{pmatrix}, \\
\label{lta_N1_2}
      \bar{\@N}^{\prime\, \mathrm{(up)}}_n(i/z_1)\sim
    \begin{pmatrix}
    \mathcal{O}(1)& e^{\zeta_1-\zeta_2}H_7E_{11}/H_1 \\
    \mathcal{O}(1) & e^{\zeta_1-\zeta_2}I_{13}E_{15}/I_6
    \end{pmatrix} ,  \\
\label{lta_N2_1}
    \bar{\@N}^{\prime\, \mathrm{(up)}}_n(\bar{z}_2)\sim
    \begin{pmatrix}
    H_7/H_1 & \mathcal{O}(e^{\zeta_1}) \\
    \mathcal{O}(e^{\zeta_1}) & \mathcal{O}(1)
    \end{pmatrix},
\\
\label{lta_N2_2}
 \bar{\@N}^{\prime\, \mathrm{(up)}}_n(i/z_2)\sim
    \begin{pmatrix}
    \mathcal{O}(1) & \mathcal{O}(e^{\zeta_1}) \\
    \mathcal{O}(e^{\zeta_1}) & I_{13}/I_6
    \end{pmatrix}.
\end{gather}
\ese
The omitted entries in the matrices above do not contribute to the asymptotic solution, either because they are exponentially small, or because they are annihilated by the norming constants in the reconstruction formula \eqref{potential}. In particular, the right column of \eqref{lta_N1_1} and the left column of \eqref{lta_N1_2} are annihilated by the zero entries of the (fundamental soliton) norming constants $\bar{\@C}_{1}(\tau)$ and $\hat{\@C}_{1}(\tau)$, respectively. The specified $\mathcal{O}(e^{\zeta_{1}})$ terms in \eqref{lta_N1_1} and \eqref{lta_N1_2} do in fact contribute to the asymptotic solution, since they multiply $\bar{\@C}_{1}(\tau)$ and $\hat{\@C}_{1}(\tau)$, which both grow at $\mathcal{O}(e^{-\zeta_{1}})$. On the other hand, the $\mathcal{O}(e^{\zeta_{1}})$ terms in \eqref{lta_N2_1} and \eqref{lta_N2_2} do not contribute, since they multiply the $\mathcal{O}(1)$ norming constants $\bar{\@C}_{2}(\tau)$ and $\hat{\@C}_{2}(\tau)$. In the above,
\bse
\begin{gather}
H_1=F_1-\frac{h_7g_6}{g_8}\Lambda e^{-2\zeta_2}, \qquad
F_1=-1+h_5\left(1-\frac{l_4h_2}{l_1f_2}\right)\Lambda e^{-2\zeta_2}, \qquad 
\Lambda=1-\frac{g_1 l_2}{g_2 l_1}, \\
H_7=\frac{h_1}{l_1}-1, \qquad
%\qquad I_1=\left(
%f_5-\frac{l_4 f_2}{l_1}+\frac{g_6 f_7}{g_8}
%\right)\Omega e^{-2\zeta_2},\qquad\Omega =1-\frac{f_1 l_2}{l_1 f_2}, \\
I_6=G_{12}-\left(\frac{f_8 f_3 l_2}{l_1^2}
+\frac{f_9 g_{12}}{g_8}
\right), \qquad G_{12}=f_{11}\left( 1-\frac{f_3 l_{10}}{f_{11} l_1}\right)e^{-2\zeta_2}-1, \\
%I_7=\frac{f_1}{l_1}-1, \qquad 
I_{13}=-1-\frac{f_8}{g_8}, \qquad %E_1=1-\frac{g_1}{l_1}, \qquad 
E_{11}=-\frac{g_6}{g_8},\qquad E_{15}=-\frac{g_{12}}{g_8}.
%E_9=\Delta \left[g_5\left(1-\frac{l_4 g_2}{l_1 g_5} \right)-\frac{g_6 g_7}{g_8} \right], \qquad
%\Delta=1-\frac{g_1 l_2}{g_2 l_1}.  %\qquad E_{13}=-\Delta \left(\frac{g_2l_{10}}{l_1}+\frac{g_{12} g_7}{g_8}\right), 
%\qquad F_{13}=\frac{h_1}{l_1}-1, \qquad F_5=h_7 \Lambda, %\qquad G_7=f_8,
\end{gather}
\ese
$l_1,\cdots,l_{12}$,  $h_1,\cdots,h_{12}$,  $f_1,\cdots,f_{12}$ and $g_1,\cdots,g_{12}$ are the coefficients in the expansions of each of Eqs.~\eqref{single_equ}, and we provide below the explicit expressions of the ones that contribute to the asymptotic expansions for the eigenfunctions above:
\bse
\begin{gather}
l_1=-\frac{4z_1^2\bar{z}_1^2}{(z_1^2-\bar{z}_1^2)^2}
\| \boldsymbol{\gamma}_1\|^2, \qquad 
l_2=-\frac{4z_2^2\bar{z}_1^2}{(z_2^2-\bar{z}_1^2)^2}
\boldsymbol{\gamma}_1^*\boldsymbol{\gamma}_2 e^{2i(b_1-b_2)n-2i(\omega_1-\omega_2)\tau}, \\
l_4=\frac{4z_1^2\bar{z}_2^2}{(\bar{z}_1^2-z_1^2)(z_1^2-\bar{z}_2^2)}W(\boldsymbol{\gamma}_2^*,\boldsymbol{\gamma}_2)\,e^{-2i(b_1-b_2)n-2i(\omega_1+\omega_2)\tau}, \\
l_{10}=
\frac{4(-1)^nz_1^2 z_2^{-2}}{(\bar{z}_1^2 - z_1^2)(z_1^2+z_2^{-2})}W(\boldsymbol{\gamma}_2,\boldsymbol{\gamma}_1)\, e^{-2i(b_1+b_2)n+2i(\omega_1+\omega_2)\tau},
\end{gather}
\begin{gather}
h_1=\frac{4z_1^2\bar{z}_1^2}{(\bar{z}_1^2-z_1^2)(z_1^2-\bar{z}_1^2)}\| \boldsymbol{\gamma}_1\|^2, \qquad
h_2=\frac{4z_2^2\bar{z}_1^2}{(\bar{z}_2^2-z_2^2)(z_2^2-\bar{z}_1^2)}\boldsymbol{\gamma}_1^* \cdot \boldsymbol{\gamma}_2\, e^{2i(b_1-b_2)n-2i(\omega_1-\omega_2)\tau}, \\
h_5=-\frac{4z_2^2\bar{z}_2^2}{(z_2^2-\bar{z}_2^2)^2}
\|\boldsymbol{\gamma}_1\|^2, \qquad
h_7=\frac{4(-1)^nz_2^2z_1^{-2}}{(\bar{z}_2^2-z_2^2)(z_2^2+z_1^{-2})}W(\boldsymbol{\gamma}_1,\boldsymbol{\gamma}_2)\,e^{-2i(b_1+b_2)n+2i(\omega_1+\omega_2)\tau}, 
\end{gather}
\begin{gather}
g_1=\frac{4z_1^2\bar{z}_1^2}{(z_1^2+z_1^{-2})(\bar{z}_1^2-z_1^2)} \|\boldsymbol{\gamma}_1\|^2, \qquad
g_2=\frac{4z_2^2\bar{z}_1^2}{(z_2^2+z_1^{-2})(\bar{z}_1^2-z_2^2)}\boldsymbol{\gamma}_1^*\cdot \boldsymbol{\gamma}_2
\, e^{2i(b_1-b_2)n-2i(\omega_1-\omega_2)\tau}, \\
%g_5=\frac{4z_2^2\bar{z}_2^2}{(z_2^2+z_1^{-2})(\bar{z}_2^2-z_2^2)} \|\boldsymbol{\gamma}_2\|^2, \qquad
g_6=\frac{4(-1)^n\bar{z}_2^2\bar{z}_1^{-2}}{(\bar{z}_1^{-2}-z_1^{-2})(\bar{z}_1^{-2}+\bar{z}_2^2)}W(\boldsymbol{\gamma}_2^*,\boldsymbol{\gamma}_1^*)\,e^{2i(b_1+b_2)n-2i(\omega_1+\omega_2)\tau}, \\
%g_7=-\frac{4(-1)^nz_2^2z_1^{-2}}{(z_2^{-2}+z_2^2)(z_2^2+z_1^{-2})}W(\boldsymbol{\gamma}_1,\boldsymbol{\gamma}_2)\,e^{-2i(b_1+b_2)n+2i(\omega_1+\omega_2)\tau}, \\
g_8=-\frac{4z_1^2\bar{z}_1^2}{(z_1^2-\bar{z}_1^2)^2}\|\boldsymbol{\gamma}_1\|^2, \qquad
g_{12}=\frac{4z_1^2\bar{z}_1^2}{(z_1^2-\bar{z}_1^2)(\bar{z}_1^2-z_2^2)}\boldsymbol{\gamma}_1^*\cdot \boldsymbol{\gamma}_2
\, e^{2i(b_1-b_2)n-2i(\omega_1-\omega_2)\tau},
\end{gather}
\begin{gather}
f_1=\frac{4z_1^2\bar{z}_1^2}{(z_1^2+z_2^{-2})(\bar{z}_1^2-z_1^2)}\|\boldsymbol{\gamma}_1\|^2, \qquad
f_2=\frac{4z_2^2\bar{z}_1^2}{(z_2^2+z_2^{-2})(\bar{z}_1^2-z_2^2)}\boldsymbol{\gamma}_1^*\cdot \boldsymbol{\gamma}_2
\, e^{2i(b_1-b_2)n-2i(\omega_1-\omega_2)\tau}, \\
f_3=\frac{4(-1)^n \bar{z}_1^2 \bar{z}_2^{-2}}{(\bar{z}_2^{-2}-z_2^{-2})(\bar{z}_1^2+\bar{z}_2^{-2})}W(\boldsymbol{\gamma}_1^*,\boldsymbol{\gamma}_2^*)\,e^{2i(b_1+b_2)n-2i(\omega_1+\omega_2)\tau}, \qquad
%f_5=\frac{4z_2^2\bar{z}_2^2}{(z_2^2+z_2^{-2})(\bar{z}_2^2-z_2^2}\|\boldsymbol{\gamma}_2\|^2, \qquad
%f_7=-\frac{4(-1)^n z_2^2 z_1^{-2}}{(z_2^{2}+z_2^{-2})(z_2^2+z_1^{-2})}W(\boldsymbol{\gamma}_1,\boldsymbol{\gamma}_2)\,e^{-2i(b_1+b_2)n+2i(\omega_1+\omega_2)\tau}, \\
f_8=\frac{4z_2^2\bar{z}_1^2}{(z_2^2-\bar{z}_1^2)(\bar{z}_1^2-z_1^2)} \|\boldsymbol{\gamma}_1\|^2, \\
f_9=\frac{4z_2^2\bar{z}_2^2}{(z_2^2-\bar{z}_2^2)(\bar{z}_2^2-z_1^2)}\boldsymbol{\gamma}_1\cdot \boldsymbol{\gamma}_2^*
\, e^{-2i(b_1-b_2)n+2i(\omega_1-\omega_2)\tau}, \qquad f_{11}=-\frac{4z_{2}^{2}\bar{z}_{2}^{2}}{(z_{2}^{2}-\bar{z}_{2}^{2})^{2}}\|\boldsymbol\gamma_{2}\|^{2}.
\end{gather}
\ese
Substituting the asymptotics~\eqref{e:LTA_2FB} into
the reconstruction formula \eqref{potential} with the explicit expressions of the involved coefficients, upon simplification we obtain for the first row:
\bse
\begin{gather}
\begin{pmatrix}
Q_{n-1}^{(1)}(\tau)\\ Q_{n-1}^{(2)}(\tau)
\end{pmatrix}
\sim -\sinh{(2a_2)}e^{2inb_2-2i\omega_2\tau}
\sech(\zeta_2-d_2^-)\@p_2^-
 \qquad \tau\to -\infty, \quad \text{fixed } \zeta_2\,,\\
\@p_2^-=
\chi\frac{\bar{z}_2^2-\bar{z}_1^2}{\bar{z}_2^2-z_1^2}\left[
\frac{\boldsymbol{\gamma}_2^*}{\|\boldsymbol{\gamma}_2\|}
+\bar{z}_1^{-2}\bar{z}_2^2\frac{\bar{z}_1^2-z_1^2}{z_1^2-\bar{z}_2^2}
\frac{\boldsymbol{\gamma}_1 \cdot \boldsymbol{\gamma}_2^*}
{\|\boldsymbol{\gamma}_1\| \|\boldsymbol{\gamma}_2\|}
\frac{\boldsymbol{\gamma}_1^*}{\|\boldsymbol{\gamma}_1 \|}-\frac{\bar z_1^2-{z}_1^2}{\bar{z}_1^2+\bar{z}_2^{-2}}
\frac{
W\left(\boldsymbol{\gamma}_1^*,\boldsymbol{\gamma}_2^* \right)}{\| \boldsymbol{\gamma}_1\|\| \boldsymbol{\gamma}_2\|}
\frac{\boldsymbol{\gamma}_1^\perp}{\| \boldsymbol{\gamma}_1\|}
\right],\\
\frac{1}{\chi^2}=
\Lambda \left[ 1+ \frac{(\bar{z}_1^2-z_1^2)(\bar{z}_2^2-z_2^2)}{(z_1^2-\bar{z}_2^2)(\bar{z}_1^2-z_2^2)}
\frac{\left| \boldsymbol{\gamma}_1 \cdot \boldsymbol{\gamma}_2^* \right|^2}{\|\boldsymbol{\gamma}_1\|^2\|\boldsymbol{\gamma}_2\|^2}
+\frac{z_2^{-2}\bar{z}_2^{-2}
(\bar{z}_1^2-z_1^2)(z_2^2-\bar{z}_2^2)}{(z_1^2+z_2^{-2})(\bar{z}_1^2+\bar{z}_2^{-2})}
\frac{\left|W \left( \boldsymbol{\gamma}_1, \boldsymbol{\gamma}_2\right)\right|^2}{\|\boldsymbol{\gamma}_1\|^2\|\boldsymbol{\gamma}_2\|^2}
\right], \\
\Lambda\equiv \frac{(\bar{z}_1^{-2}-\bar{z}_2^{-2})(z_2^{-2}-z_1^{-2})}{(z_1^{-2}-\bar{z}_2^{-2})(z_2^{-2}-\bar{z}_1^{-2})}\,, \qquad d_2^-=d_2^+-\log \chi\,, \qquad 
d_2^+=\frac{\log{\|\boldsymbol{\gamma}_2\|}}{\sinh{(2a_2)}}\,.
\end{gather}
\ese
The above expression is a 1-fundamental soliton solution as in \eqref{e:1FS}, with a polarization vector $\@p_2^-$.

It can be seen that this is consistent with \eqref{u2-} and \eqref{v2-} by noting that
\bse
\begin{gather}
\frac{W(\boldsymbol\gamma_{1}^{*},\boldsymbol\gamma_{2}^{*})}{\|\boldsymbol\gamma_{1}\|\|\boldsymbol\gamma_{2}\|}\frac{\boldsymbol\gamma_{1}^{\perp}}{\|\boldsymbol\gamma_{1}\|}=\frac{\boldsymbol\gamma_{2}^{*}}{\|\boldsymbol\gamma_{2}\|}-\frac{(\boldsymbol\gamma_{1}\cdot\boldsymbol\gamma_{2}^{*})}{\|\boldsymbol\gamma_{1}\|\|\boldsymbol\gamma_{2}\|}\frac{\boldsymbol\gamma_{1}^{*}}{\|\boldsymbol\gamma_{1}\|}\,,\\
\frac{|W(\boldsymbol\gamma_{1},\boldsymbol\gamma_{2})|^{2}}{\|\boldsymbol\gamma_{1}\|^{2}\|\boldsymbol\gamma_{2}\|^{2}}=1-\frac{|\boldsymbol\gamma_{1}\cdot\boldsymbol\gamma_{2}^{*}|^{2}}{\|\boldsymbol\gamma_{1}\|^{2}\|\boldsymbol\gamma_{2}\|^{2}}\,.
\end{gather}
\ese
Substituting these into the above, after simplification we find
\bse
\begin{gather}
    \@p_{2}^{-}=\chi\frac{(\bar{z}_{2}^{2}-\bar{z}_{1}^{2})(\bar{z}_{2}^{-2}+z_{1}^{2})}{(\bar{z}_{2}^{2}-z_{1}^{2})(\bar{z}_{2}^{-2}+\bar{z}_{1}^{2})}\left[\frac{\boldsymbol\gamma_{2}^{*}}{\|\boldsymbol\gamma_{2}\|}+\frac{(\bar{z}_{1}^{2}-z_{1}^{2})(\bar{z}_{1}^{-2}+z_{1}^{2})}{(z_{1}^{2}-\bar{z}_{2}^{2})(\bar{z}_{2}^{-2}+z_{1}^{2})}\frac{(\boldsymbol\gamma_{1}\cdot\boldsymbol\gamma_{2}^{*})}{\|\boldsymbol\gamma_{1}\|\|\boldsymbol\gamma_{2}\|}\frac{\boldsymbol\gamma_{1}^{*}}{\|\boldsymbol\gamma_{1}\|}\right]\,,\\
    \label{chi_inv}
    \frac{1}{\chi^{2}}=\bigg\vert\frac{(\bar{z}_{2}^{2}-\bar{z}_{1}^{2})({z}_{1}^{2}+\bar z_{2}^{-2})}{(\bar{z}_{2}^{2}-z_{1}^{2})(\bar z_{1}^{2}+\bar z_{2}^{-2})}\bigg\vert^{2}\Bigg[1+\frac{(\bar{z}_{1}^{2}-z_{1}^{2})(z_{2}^{2}-\bar{z}_{2}^{2})(z_{1}^{2}+\bar{z}_{1}^{-2})(\bar{z}_{2}^{-2}+z_{2}^{2})}{(z_{2}^{2}-\bar{z}_{1}^{2})(z_{1}^{2}-\bar{z}_{2}^{2})(\bar{z}_{1}^{-2}+z_{2}^{2})(z_{1}^{2}+\bar{z}_{2}^{-2})}\frac{|\boldsymbol\gamma_{1}\cdot\boldsymbol\gamma_{2}^{*}|^{2}}{\|\boldsymbol\gamma_{1}\|^{2}\|\boldsymbol\gamma_{2}\|^{2}}\Bigg]\,.
\end{gather}
\ese
The expression for $\@p_{2}^{-}$ agrees exactly with \eqref{u2-} and \eqref{v2-} with $\@v_{1}^{-}=\@v_{2}^{+}=(1,0)^{T}$. Furthermore, in the notation of Section 4, from
\begin{equation}
    \frac{1}{\chi_{\gamma}^{2}}=\frac{\bar{z}_{2}^{2}}{z_{2}^{2}}\frac{\|\boldsymbol\gamma_{2}^{-}\|^{2}}{\|\boldsymbol\gamma_{2}^{+}\|^{2}}\,,
\end{equation}
a direct calculation shows that
\begin{equation}
    \frac{1}{\chi_{\gamma}^{2}}=\frac{\bar{z}_{2}^{2}}{z_{2}^{2}}\bigg\vert\frac{{z}_{1}^{-2}+\bar z_{2}^{2}}{\bar z_{1}^{-2}+\bar z_{2}^{2}}\bigg\vert^{2}\Bigg[1+\frac{(\bar{z}_{1}^{2}-z_{1}^{2})(z_{2}^{2}-\bar{z}_{2}^{2})(z_{1}^{2}+\bar{z}_{1}^{-2})(\bar{z}_{2}^{-2}+z_{2}^{2})}{(z_{2}^{2}-\bar{z}_{1}^{2})(z_{1}^{2}-\bar{z}_{2}^{2})(\bar{z}_{1}^{-2}+z_{2}^{2})(z_{1}^{2}+\bar{z}_{2}^{-2})}\frac{|\boldsymbol\gamma_{1}\cdot\boldsymbol\gamma_{2}^{*}|^{2}}{\|\boldsymbol\gamma_{1}\|^{2}\|\boldsymbol\gamma_{2}\|^{2}}\Bigg]\,.
\end{equation}
Similarly, with $\@v_{1}^{-}=\@v_{2}^{+}=(1,0)^{T}$ we have
\begin{equation}
    \frac{1}{\chi_{\delta}^{2}}=\frac{z_{2}^{2}}{\bar{z}_{2}^{2}}\frac{z_{1}^{4}}{\bar{z}_{1}^{4}}\bigg\vert\frac{\bar{z}_{2}^{2}-\bar{z}_{1}^{2}}{\bar z_{2}^{2}-{z}_{1}^{2}}\bigg\vert^{2}\,.
\end{equation}
Putting them together and simplifying the factor in front gives
\begin{equation}
    \frac{1}{\chi^{2}}=\frac{1}{\chi_{\gamma}^{2}\chi_{\delta}^{2}}=\bigg\vert\frac{(\bar{z}_{2}^{2}-\bar{z}_{1}^{2})({z}_{1}^{2}+\bar z_{2}^{-2})}{(\bar{z}_{2}^{2}-z_{1}^{2})(\bar z_{1}^{2}+\bar z_{2}^{-2})}\bigg\vert^{2}\Bigg[1+\frac{(\bar{z}_{1}^{2}-z_{1}^{2})(z_{2}^{2}-\bar{z}_{2}^{2})(z_{1}^{2}+\bar{z}_{1}^{-2})(\bar{z}_{2}^{-2}+z_{2}^{2})}{(z_{2}^{2}-\bar{z}_{1}^{2})(z_{1}^{2}-\bar{z}_{2}^{2})(\bar{z}_{1}^{-2}+z_{2}^{2})(z_{1}^{2}+\bar{z}_{2}^{-2})}\frac{|\boldsymbol\gamma_{1}\cdot\boldsymbol\gamma_{2}^{*}|^{2}}{\|\boldsymbol\gamma_{1}\|^{2}\|\boldsymbol\gamma_{2}\|^{2}}\Bigg]\,,
\end{equation}
which is indeed the same as \eqref{chi_inv}.

%\numberwithin{equation}{section}
%\newpage
%{\color{blue}References not cited in text:}

\end{document}